\documentclass[onecolumn,showpacs,preprintnumbers,amsmath,amssymb,superscriptaddress]{revtex4-2}
\usepackage{graphicx}% Include figure files
\usepackage{dcolumn}% Align table columns on decimal point
\usepackage{bm}% bold math
\usepackage{tabularx}
\usepackage{makecell}
\usepackage{braket}
\usepackage{lipsum}
\usepackage{ulem}
\usepackage{amsmath,amssymb}
\newcolumntype{M}{>{\centering\arraybackslash}m{1.85cm}}
\usepackage[export]{adjustbox}
\usepackage{longtable}
\usepackage{float}
\usepackage{xcolor}
\usepackage{footnote}
\usepackage{color}   %May be necessary if you want to color links
\usepackage{hyperref}
\hypersetup{
	colorlinks=true, %set true if you want colored links
	linktoc=all,     %set to all if you want both sections and subsections linked
	linkcolor=blue,  %choose some color if you want links to stand out
}

\makeatletter
\newcommand{\colorcaption}[2][]{%
	\begingroup%
	\renewcommand{\@caption@fignum@sep}{ (Color online). }%
	\caption[#1]{#2}%
	\endgroup%
}
\makeatother
\newcommand{\orcid}[1]{\href{https://orcid.org/#1}{\hskip2pt\includegraphics[width=9pt]{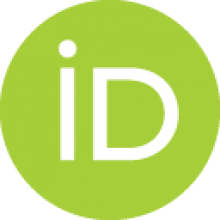}}}

\begin{document}

\title{Shell model study of isobaric analog states for $T_z= \pm 2$  nuclei using isospin non-conserving interactions}

\author{Sakshi Shukla}
\address{Department of Physics, Indian Institute of Technology Roorkee, Roorkee
	247 667, India}
	\author{Praveen C. Srivastava\footnote{Contact author: praveen.srivastava@ph.iitr.ac.in}}
\address{Department of Physics, Indian Institute of Technology Roorkee, Roorkee
	247 667, India}
   \author{ Kazunari Kaneko }
\address{Department of Physics, Kyushu Sangyo University, Fukuoka 813-8503, Japan}

\vspace{10pt}
%\begin{indented}
%\item[]August 2017
%\end{indented}

\begin{abstract}
In order to comprehend the process underlying mirror energy differences in mirror pairs, we have performed shell-model calculations for $T_z= \pm 2$ $sd$-shell nuclei in the mass range $A$= 20 to 36 and neutron numbers varying from $N$= 8 to 20.
Isospin-symmetry breaking (ISB) is responsible for the mirror energy difference of excited states.  We have investigated the {\color{black}isospin non-conserving} interactions: USDC and USDCm to explore the low-lying energy spectra, mirror energy differences, isoscalar ($M_0$), isovector ($M_1$) matrix elements, \textit{E2} transition probability, magnetic ($\mu$), and quadrupole moments ($Q$) of mirror-pair and compared them with their available experimental data. The impact of single-particle states on weakly bound and unbound nuclear states are investigated, especially those of the $s$-wave. We have also analyzed single proton/neutron separation energies and proton/neutron occupancy for ($T_z$=-2)/($T_z$=+2) $sd$-shell nuclei. 
%In low-lying states, USDC provides a more accurate prediction of isospin mixing, whereas USDCm is more effective for mirror energy differences.
\end{abstract}

%
% Uncomment for keywords
%\vspace{2pc}
%\noindent{\it Keywords}: XXXXXX, YYYYYYYY, ZZZZZZZZZ
%
% Uncomment for Submitted to journal title message
%\submitto{\JPA}
%
% Uncomment if a separate title page is required

% 
% For two-column output uncomment the next line and choose [10pt] rather than [12pt] in the \documentclass declaration
%\ioptwocol
%

\maketitle
%\maketitle
\section{Introduction}
\label{sect 1}
%According to Bohr \cite{bohr}, the nuclear force holds protons and neutrons together in nuclei. 
{\color{black}Proton and neutron are two distinct charge states of the same particle, they  differ by their isospin projection ($t_z$) \cite{Heisenberg, wigner}}.
%The isospin projection ($t_z$) is the only difference between the proton and neutron, which are two distinct states of the same particle . 
%According to Bohr \cite{bohr}, the nuclear force holds protons and neutrons together in nuclei. 
%The isospin projection ($t_z$) is the only difference between the proton and neutron, which are two distinct states of the same particle \cite{Heisenberg, wigner}. 
Mirror nuclei are pairs of nuclei in which the proton and neutron numbers are exchanged, offering unique opportunities to investigate isospin symmetry and its breaking phenomena. Isospin symmetry is known as the degeneracy of energy levels of mirror nuclei. The mass difference between up and down quarks and the effects of electromagnetic interaction cause a slight breaking of the symmetry. %This phenomenon is referred as charge independence breaking (CIB) or isospin symmetry breaking.
This phenomenon is referred to as isospin symmetry breaking (ISB).
{\color{black} The algebraic approach to explain 
the ISB is given in Ref. \cite{piet}.}
{\color{black}The concept of isospin-symmetry breaking can be viewed as a dynamical symmetry breaking that does not admix isospin $T$ (but splits states with the same $T$ but different $T_Z$) if the off-diagonal isospin mixing matrix elements of the Coulomb interaction and of nuclear origin are neglected, but not in the
situation when isospin mixing matrix elements of both interactions are considered.} Although the Coulomb force is the primary cause of isospin-symmetry breakdown, theoretical study has indicated the charge dependency of the nuclear force may also play a significant role \cite{suzuki,lenzi,kaneko_1}.
Understanding ISB can help us to test nuclear models and understand the fundamental forces inside the nucleus \cite{Nadya,ISB_rev,chandan_ISB,kaneko_2}. One of the significant observations in investigating ISB is the mirror energy difference (MED) \cite{bentley_2,gadea,Stroberg}. MED for $sd$-shell nuclei have been evaluated in these Refs. \cite{MED_A20,MED_sd1,MED_sd2,Sakshi_ISB}.  MED is the difference between the excitation energies of mirror pairs, given as
\begin{equation*}
    MED(A, T, J^{\pi})= E_{ex}(A, T, J^{\pi}, T_{z_<})-E_{ex}(A, T, J^{\pi}, T_{z_>}),
\end{equation*}
{\color{black} here, $E_{ex}$ is the excitation energy, $T_{z<}$ stands for the isospin of proton-rich nucleus and $T_{z>}$ stands for the neutron-rich nucleus.}
Several theoretical models are available to investigate MED values in $sd$-shell nuclei such as \textit{ab initio} approach in which experimental energy levels in $sd$-shell nuclei were used to constrain linear combinations of TBMEs \cite{Stroberg}. Another well-known $sd$-shell shell model (SM) Hamiltonians, such as USD \cite{Wildenthal, Brown}, USDA \cite{Richter}, and USDB \cite{Richter} interactions in which analytic Coulomb interaction is combined with Miller-Spencer short-range correlations {\color{black}(Jastrow-type correlation of the form $V(r) = (1+f(r))/r$)}  and an effective isotensor interaction, which helps us to better understand exotic structures of nuclei and ISB.
Two new USD-type interactions are developed by including {\color{black}isospin non-conserving} and Coulomb part: USDC and USDI \cite{magilligan}.
The USDC interaction is built by using the same renormalized G-matrix which was used in the previous USD-type Hamiltonians, while the USDI interaction is obtained from the in-medium similarity renormalization group. 
{\color{black} There are two additional interactions are
USDCm and USDIm in which the Coulomb TBME are further
constrained  \cite{magilligan}.}
%There are two additional slightly modified interactionsThere are two additional slightly modified interactions: USDCm and USDIm. 
In mirror-pair nuclei, when the proton-rich nucleus is near the proton drip-line, it exhibits a weakly bounded nature such as $^{30}$Cl and $^{34}$K, whereas the corresponding neutron-rich counterpart shows deeply bounded behavior such as $^{30}$Al and $^{34}$P. In certain cases {\color{black} for  e.g., in} those mirror states where angular momentum $l$=0, they exhibit large MED values with considerable ISB; this phenomenon is known as the Thomas-Ehrman (TE) shift \cite{TE_1,TE_2}. In Ref. \cite{MED_22Al} a significant TE shift is observed due to mirror energy differences between the mirror-pairs $^{22}$Al and $^{22}$F, which were explained by the continuum effect.
 
 There are some recent experiments done for the $sd$-shell nuclei, such as for the $^{22}$Al, $T_z = -2$ nuclei, a few experimental states $4^+_1$, $1^+_1$, and $1^+_2$ are observed at 0, 905 (403), and 2145 (403) in Ref. \cite{lee}, respectively, and in Ref. \cite{22Al_expt} $0^+_1$ isobaric analog state is observed at 8829 (406) keV excitation energy which is followed by two-proton emission. In Ref. \cite{isomer_26P} for the A=26 ($T_z$ = $\pm$2) mirror pair an 
 isomeric state is observed by using $\gamma$-spectroscopy at 0.164 MeV excitation energy and with a half-life equal to 120(9) ns for $^{26}$P
 (T$_z$= -2), although its spin and parity are not known. For the $^{26}$Na (T$_z$= +2) an isomeric state at 0.082 MeV excitation energy with a revised 
 half-life value equal to 4.35 (16) $\mu s$ \cite{isomer_26P} is observed. In Ref. \cite{34P_36S} yrast and non-yrast states are observed in $T_z=+2$ nuclei, $^{34}$P, and $^{36}$S using transfer/deep-inelastic processes. Observed experimental levels are compared with the theoretical result obtained from the $sdpf$ calculation, which indicates that even for the low-lying positive-parity states, it is crucial to include the orbitals from the higher $pf$ shell. However, in the present calculation we have used only the $sd$-shell interaction. Earlier in Ref. \cite{36Ca_states}, the first excited state $2^+$ in 
 $^{36}$Ca is observed at 3.015 (16) MeV excitation energy, which lies 276 keV lower than its corresponding mirror pair $^{36}$S, which 
provides experimental MED value -276 (16) keV. In a recent experiment performed by Lalanne \textit{et al.} \cite{36Ca_expt}, the structure of 
$^{36}$Ca was investigated using Coulomb excitation following which some new states are observed such as $0^+_2$, $1^+_1$, $2^+_2$, and $0^+_3$  states at 2.83 (13), 4.24 (4), 4.71 (9), and 4.83 (17) MeV excitation energies, respectively. Similar shifts are obtained in the energies of $1^+_1$ and $2^+_1$ states as in the case of its mirror pair $^{36}$S (T$_z$=+2). Among these newly observed states, $0^+_2$ state is an intruder state, whereas in the case of its mirror pair $^{36}$S, the second $0^+$ state is not an intruder state. 
{\color{black} The mirror energy difference has important implications in nuclear astrophysics, the astrophysical reaction rate for the study of
the $T = 1/2$ mirror nuclei, $^{31}$S and $^{31}$P is reported in Ref. \cite{MED_AP}. Because of the significance in electromagnetic spin-orbit splitting, the large mirror energy differences between the  $9/2_1^-$ and $13/2_1^-$ states were observed in comparison to the $7/2_1^-$ and $11/2_1^-$ \cite{MED_AP1}. In Ref. \cite{panu}, a large difference in the B(E2; $5/2_1^+$ $\rightarrow$ $1/2_1^+$) value between $T = 3/2$, A = 21 mirror nuclei has been observed.}

%Masses are key inputs for many nuclear astrophysics processes. 
%for e.g., the precise calculation of mirror energy difference between $^{43}$V and $^{43}$Ca is very crucial to predict  thermonuclear rate of $^{42}$Ti($p$,$\gamma$)$^{43}$V and its astrophysical implication in the $rp$ process \cite{MED_AP}.}

Earlier, we have explored $T_z$=$\pm$3/2 $sd$-shell mirror pairs \cite{Sakshi_ISB} using newly developed USDC and USDCm effective interactions, the aim of the present work is to examine the $T_z= \pm 2$ mirror pair of nuclei. By using these interactions, we have computed the energy spectra, mirror energy difference, occupancy, $B(E2)$ transitions, single neutron/proton separation energies, quadrupole and magnetic moments of these nuclei.

\section{Model Space and Isospin Non-Conserving Hamiltonians}

The shell-model Hamiltonian can be mathematically represented as a sum of one- and two-body
 operators as follows,
\begin{equation}
H=\sum_{\alpha}\varepsilon_{\alpha}{\hat N}_{\alpha}+\frac{1}{4}\sum_{\alpha\beta\delta\gamma JT}\langle j_{\alpha}j_{\beta}|V|j_{\gamma}j_{\delta}\rangle_{JT}A^{\dag}_{JT;j_{\alpha}j_{\beta}} \times 
 A_{JT;j_{\delta}j_{\gamma}},
\end{equation}
here, the single-particle orbitals are denoted by $\alpha=\{nljt\}$ and the associated single-particle energies are denoted by $\varepsilon_{\alpha}$. $\hat{N}_{\alpha}=\sum_{j_z,t_z}a_{\alpha,j_z,t_z}^{\dag}a_{\alpha,j_z,t_z} $ represents the particle number operator. Two-body matrix elements $\langle j_{\alpha}j_{\beta}|V|j_{\gamma}j_{\delta}\rangle_{JT}$ are coupled to the spin $J$ and isospin $T$. The fermion creation and annihilation operators are denoted by $A_{JT}^{\dag}$ and $A_{JT}$, respectively.

The Hamiltonian of USDC and USDCm effective interactions \cite{magilligan,brown_note} used in the present work which includes three components, $H$ = $H_0$ + $H_{INC}$ + $H_{C}$, where $H_0$ is the isospin conserving part which assumes charge independence, meaning that protons and neutrons interact identically via the strong force. Therefore, $V_{pp}$ = $V_{nn}$ = $V_{pn} (T = 1)$. The second part is the $H_{INC}$, i.e., {\color{black}isospin non-conserving part of nuclear interaction} and $H_{C}$ is the Coulomb part, {\color{black} then both interactions allow us to introduce the isospin non-conservation.}
%then the single-particle energy and TBMEs of $V_{pp}$, and $V_{nn}$ parts of the interaction become different. 
It is well known that energy splitting in an isobaric multiplet cannot be entirely explained by the Coulomb interaction alone; this discrepancy is known as the Nolen-Schiffer anomaly \cite{NS}. Although the Coulomb force is the primary cause of ISB, to explain these ISB effects in atomic nuclei, the charge-dependent nuclear force must be included with the Coulomb interaction. To account for the charge-dependent effects, the $T=1$ proton-neutron interaction term $V_{pn}$ of the USDC and USDCm SM Hamiltonians has been increased by 2.2\% and 0.8\%, respectively.

{\color{black}The findings of this study are shown in Tables 1-3 and Figs. 1-13. In Table \ref{T_3_2}, the theoretical $B(E2)$ values are shown along with their available experimental data.
For the $B(E2)$ calculation, we have taken effective charges $e_p$=1.36e and $e_n$=0.45e \cite{E2_sd}.} To make a more refined test of the used interaction for $B(E2)$ values, we have also studied
isocalar ($M_0$) and isovector ($M_1$) matrix element corresponding to the $E2$ transition. Where, $M_0$ and $M_1$ is given as,
\begin{equation}
M_0=\frac{\sqrt{B(E2;T_z<0)}+\sqrt{B(E2;T_z>0)}}{2},
\end{equation}
and
\begin{equation}
M_1=\left|\frac{\sqrt{B(E2;T_z<0)}-\sqrt{B(E2;T_z>0)}}{\Delta T_z}\right|.
\end{equation}
Here, $B(E2;T_z<0)$ and $B(E2;T_z>0)$ correspond to $B(E2)$ values for the proton- and neutron-rich mirror nuclei, respectively. $\Delta T_z $ gives the difference of isospin values in mirror pairs. In the present case for $T_Z=\pm 2$, the value of $\Delta T_z $ is equal to 4. For occupancy, {\color{black}we have plotted average proton and neutron occupancies for the proton and neutron rich mirror nuclei. Additionally, the proton and neutron thresholds are indicated in the energy spectra.}

%%%%%%%%%%%%%%%%%%%%%%%%%%%%%%%%%%%%%%%%%%%%%%%%%%%%%%%%%%%%%%%%%%%%%%%%%%%%
\section{Results and Discussions:}

\begin{figure*}
\hspace{-1.0cm}
\includegraphics[width=8.750cm,height=10cm]{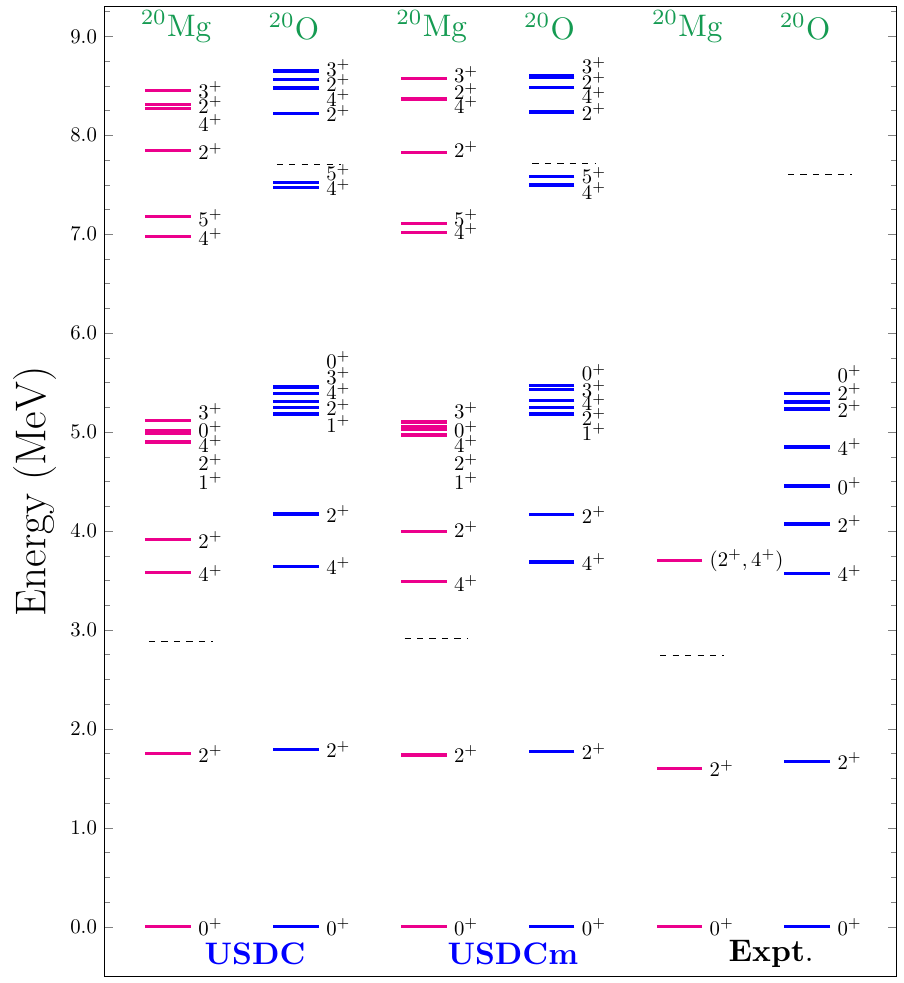}
\includegraphics[width=8.750cm,height=10cm]{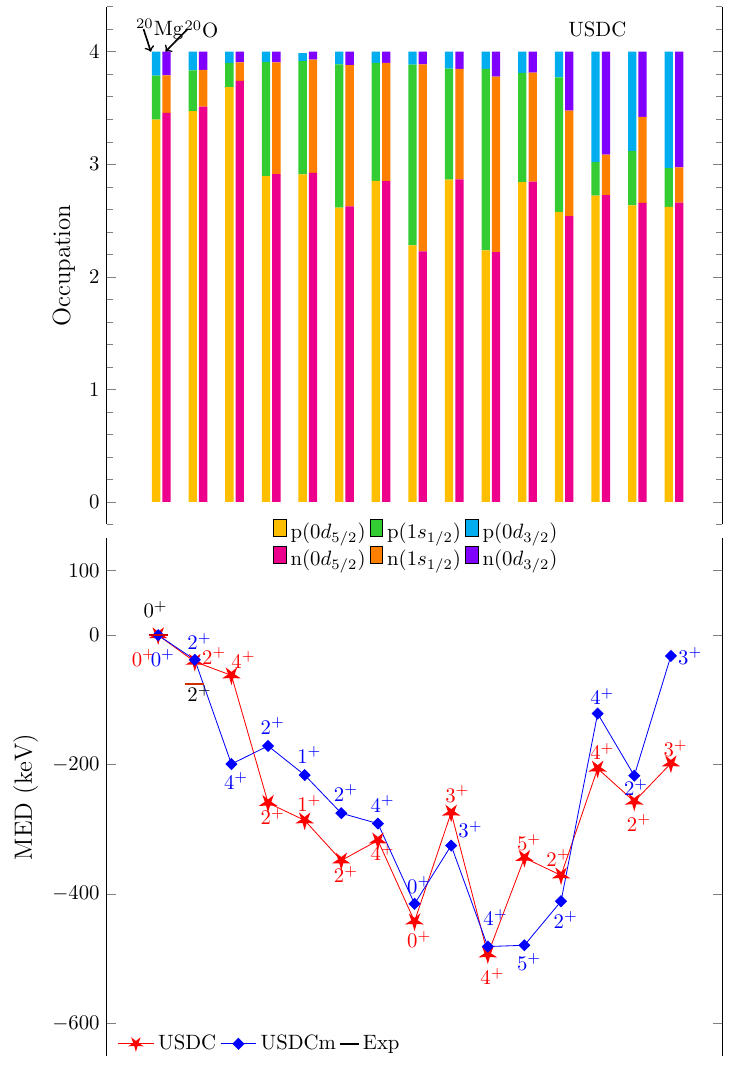}
\caption{\label{20_usdc} Comparison between the calculated and experimental \cite{NNDC} energy levels (solid lines) and the proton and neutron thresholds (dashed lines)(left), mirror energy differences for low-lying states {\color{black} and proton and neutron
occupancies of single-particle orbits for isobaric analog states of $^{20}$Mg and $^{20}$O (right).}}
\end{figure*}

\begin{table*}        
\begin{center}
%\begin{threeparttable}
\caption{Comparison between the calculated and experimental $B(E2)$ $\downarrow$ [e$^2$fm$^4$] values for $T_z$=$\pm 2$ nuclei for the effective charges $e_p$=1.36e and $e_n$=0.45e \cite{E2_sd}.}
\label{T_3_2}
%\begin{ruledtabular}
\begin{tabular}{ccccccccc}
%\hline
% & & &  \multicolumn{5}{c}
% {B(E2)$\downarrow$ %[e$^2$fm$^4$]}& \\
 \hline \\[-7pt]
Isotope 		& 	$J^\pi_i$ 			& 	$J^\pi_f$ 			& 	Expt. 		& 	VS-IMSRG 	&USDB&USDC& 	USDCm 	& 	Ref$._{Expt}$ 			\\[+2pt]
\hline
$^{20}$Mg 	&$2^+_1$ &$0^+_1$&	35.4 (64)		& 	26.3		&37.6	&	38.7&38.3	&\cite{NNDC} 		\\[+1pt]
$^{20}$O &$2^+_1$ &$0^+_1$&	5.8 (2)		& 	0.9		&4.1	&4.1	&4.1&	\cite{NNDC} 		\\[+1pt]
$^{22}$Al & $3^+_1$  &$4^+_1$ &NA & -&66.5 &67.2  & 67.1&\\[+1pt]
$^{22}$F &  $3^+_1$ & $4^+_1$ &NA & -&37.3 & 37.7 & 37.8&\\[+1pt]
$^{24}$Si 	 &$2^+_1$ &$0^+_1$&	19.0	(57)		& 	41.4&47.3		&	50.0&	48.5&	\cite{NNDC} 		\\[+1pt]
$^{24}$Ne	 &$2^+_1$ &$0^+_1$&	28.0	(66)		&	13.8& 40.6		&	41.3&	41.3	&	\cite{NNDC} 		\\[+1pt]
$^{26}$P & $1^+_1$  &$3^+_1$ &NA &- &27.9 &29.6  & 27.3&\\[+1pt]
$^{26}$Na & $1^+_1$  &$3^+_1$ & 16.5 (41)&- &35.0 &36.6  & 36.2&\cite{NNDC}\\[+1pt]
$^{28}$S 	 &$2^+_1$ &$0^+_1$&	36.2	(60) 	&	45.1& 50.6		&	50.4	&49.5	&	\cite{Togano} 		\\[+1pt]
$^{28}$Mg	 &$2^+_1$ &$0^+_1$&	67.7	(61)	&	28.4 & 63.5		&	63.8 &63.3	&	\cite{NNDC} 		\\[+1pt]
$^{30}$Cl & $2^+_1$  &$3^+_1$ & NA&- &24.6 &23.8  & 24.3&\\[+1pt]
$^{30}$Al & $2^+_1$  &$3^+_1$ & NA&- &29.1 &26.3  & 27.4&\\[+1pt]
$^{32}$Ar 	&$2^+_1$ &$0^+_1$	&	53.2	(136) &	37.1 & 53.5		&	53.1&	53.2&	\cite{32Ar_expt} 		\\[+1pt]
$^{32}$Si	&$2^+_1$ &$0^+_1$	& 	32.0	(91) & 	21.3&44.5		&	45.8	&44.0&	\cite{NNDC} 		\\[+1pt]
$^{34}$K & $2^+_1$  &$1^+_1$ &NA & -& 2.7& 2.8& 2.7  &\\[+1pt]
$^{34}$P & $2^+_1$  &$1^+_1$ &NA & -& 0.4& 0.4 &0.3 &\\[+1pt] % there is experimental data regarding this but the amount is very high 327$_{+850}^{-327}$
$^{36}$Ca & $2^+_1$  &$0^+_1$ & 26.2 (40)& - &2.4 &2.4  & 2.4&\cite{be2_36Ca}\\[+1pt]
$^{36}$S & $2^+_1$  &$0^+_1$ &20.0 (17) & -&21.6 &21.1  & 20.6&\cite{NNDC}\\[+1pt]
\hline
\end{tabular}
%\end{ruledtabular}
\vspace{-10pt}
\end{center}
%\end{threeparttable}
\end{table*}

\begin{figure}
\centerline{\includegraphics[width=0.5\linewidth]{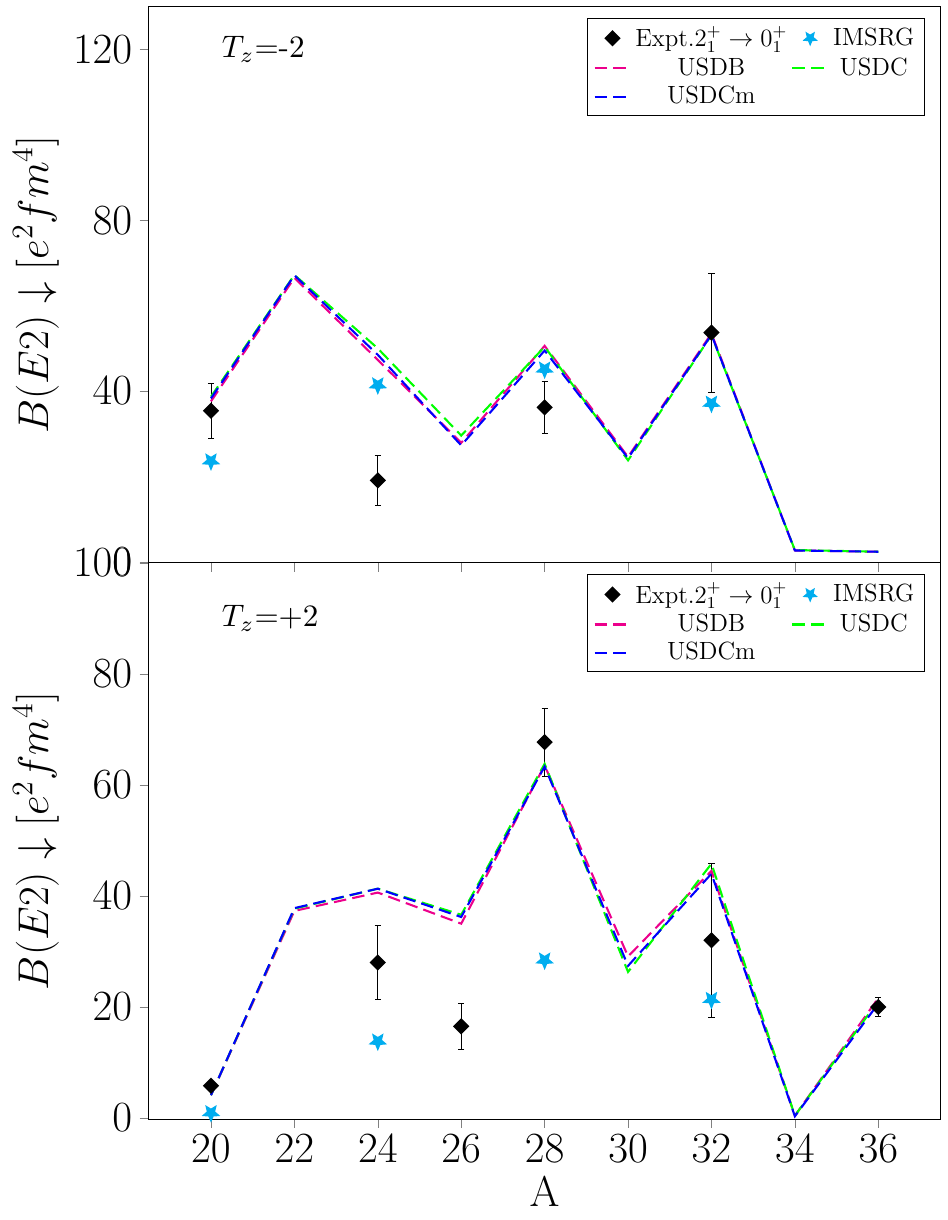}}
\caption{$B(E2)\downarrow$ values for (top) $T_z=-{2}$ and (bottom) $T_z=+{2}$ nuclei. We have taken result of VS-IMSRG interaction from the Ref. \cite{E2_sd}.}
\label{fig:be2}
\end{figure}

\begin{table*}        
\begin{center}
%\begin{threeparttable}
\caption{Single proton separation energy ($s_{1p})$ for $T_z$=$-2$, and single neutron separation energy ($s_{1n}$) for $T_z$=$+2$ (using USDC interaction) nuclei in comparison with the experimental data \cite{NNDC}.}
\label{sep}
%\begin{ruledtabular}
\begin{tabular}{ccc|ccc}
% & & &  \multicolumn{5}{c}{B(E2)$\downarrow$ [e$^2$fm$^4$]}& \\ \hline \\[-7pt]
\hline
Isotope &Expt.		& 	$s_{1p}$  & Isotope &Expt. & $s_{1n}$				\\[+2pt]
 \hline \\[-10pt]
$^{20}$Mg& 2.741(11)& 2.884&$^{20}$O &7.608(3) &7.704\\[+1pt]
$^{22}$Al&0.001(401)&0.040 &$^{22}$F &5.230(13) &5.138\\[+1pt]
$^{24}$Si&3.292(19) &3.491 &$^{24}$Ne&8.869(1) &8.916 \\[+1pt]
$^{26}$P&0.140(2) & -0.152&$^{26}$Na&5.574(4) & 5.501\\[+1pt]
$^{28}$S&2.490(16) &2.586 &$^{28}$Mg &8.503(2) &8.649\\[+1pt]
$^{30}$Cl&-0.480(20) &-0.688 &$^{30}$Al & 5.727(2)&5.609\\[+1pt]
$^{32}$Ar&2.455(5) &2.640 &$^{32}$Si &9.200(0) &9.362\\[+1pt]
$^{34}$K& -0.900(298)&-0.744 &$^{34}$P &6.282(1) &6.198\\[+1pt]
$^{36}$Ca&2.570(4) &2.617 &$^{36}$S &9.889(0) &9.964\\[+1pt]
\hline
\end{tabular}
%\end{ruledtabular}
\vspace{-10pt}
\end{center}
%\end{threeparttable}
\end{table*}

{\bf $^{20}$Mg/$^{20}$O:} Fig. \ref{20_usdc} shows the comparisons between the experimental data \cite{NNDC} and results obtained using USDC and USDCm interactions for the A = 20 mirror pair nuclei $^{20}$Mg/$^{20}$O,
resulting in the $0^+$ ground state for the $^{20}$Mg and $^{20}$O, which supports the experimental assignment. Experimental states up to 4.5 MeV are reproduced very well for $^{20}$O. Experimentally tentative states ($2^+, 4^+$) are available at 3.70 MeV, theoretically $4^+_1$ state is obtained at 3.582 and 3.487 MeV excitation energies using the USDC and USDCm interactions, respectively. Whereas the $2^+_2$ state lies at 3.914 and 3.993 MeV excitation energies using USDC and USDCm interactions, respectively, this indicates that the tentative state at 3.70 MeV is much closer to $4^+_1$ state, therefore this tentative state might belong to the $4^+_1$ state in $^{20}$Mg. All the MED values obtained using USDC and USDCm interactions are negative because the excitation energies of proton-rich nuclear states are higher than their neutron-rich mirror states. In this case, we get a large MED of -493 and -442 keV corresponding to $4^+_3$ and $0^+_2$ states using USDC interaction, respectively, due to the higher occupancy of the $s_{1/2}$ orbital, which suggests there might be a TE shift in these states. The USDCm interaction provides a large MED for the $4^+_3$ and $5^+_1$ states. The configuration of $0^+_1$, $2^+_1$, and $4^+_1$ states in $^{20}$Mg/$^{20}$O are $\pi(d_{5/2}^4)$/$\nu(d_{5/2}^4)$ i.e. obtained from one proton/neutron pair breaking in the $d_{5/2}$ orbital. The $1^+_1$, $3^+_1$, and $5^+_1$ states are obtained by four proton/neutron $d_{5/2}^3s_{1/2}$ configurations in $^{20}$Mg/$^{20}$O i.e. by one proton/neutron pair breaking in $d_{5/2}$ i.e. $(d_{5/2}^2)_{0^+,2^+,4^+}\otimes d_{5/2}$, and one unpaired proton/neutron in the $s_{1/2}$ orbital. In both cases for $^{20}$Mg and $^{20}$O the B(E2;$2^+_1 \rightarrow0^+_1$) values are {\color{black} reasonably} reproduced by $sd$- shell interactions: USDB, USDC, and USDCm than VS-IMSRG as we can see in Table \ref{T_3_2}. The calculated isoscalar and isovector matrix elements for the $2^+_1 \rightarrow0^+_1$ $E2$ transition also shows good agreement with their experimental values as we can see in Fig. \ref{fig:be2}. The calculated deformation parameter ($\beta_2$) using USDC and USDCm interactions for the $2^+_1$ state in $^{20}$Mg are 0.46 and 0.45, respectively. The corresponding experimental value is 0.44 \cite{NNDC}. The magnetic moments for the $2^+_1$ state in $^{20}$O are -0.707 and -0.711 $\mu_N$ using USDC and USDCm interactions, respectively, which is in good agreement with its experimental value i.e., -0.70 (3) $\mu_N$. The calculated and experimental single neutron separation energies are higher for the $^{20}$O in comparison to the single proton separation energy in $^{20}$Mg as shown in Table \ref{sep}, due to greater proton-proton repulsion in $^{20}$Mg. Therefore, $^{20}$O is more deeply bound than $^{20}$Mg.

\begin{figure*}
\hspace{-1.0cm}
\includegraphics[width=8.750cm,height=10cm]{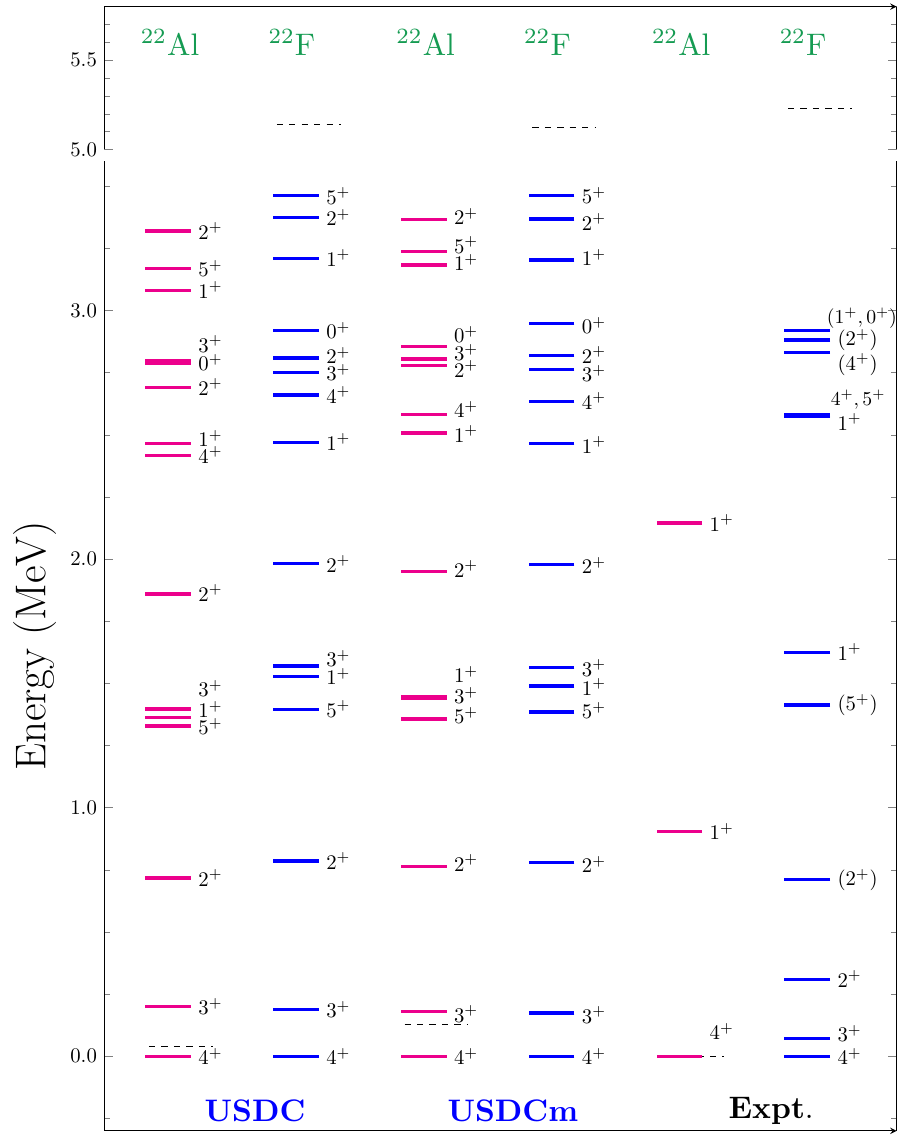}
\includegraphics[width=8.750cm,height=10cm]{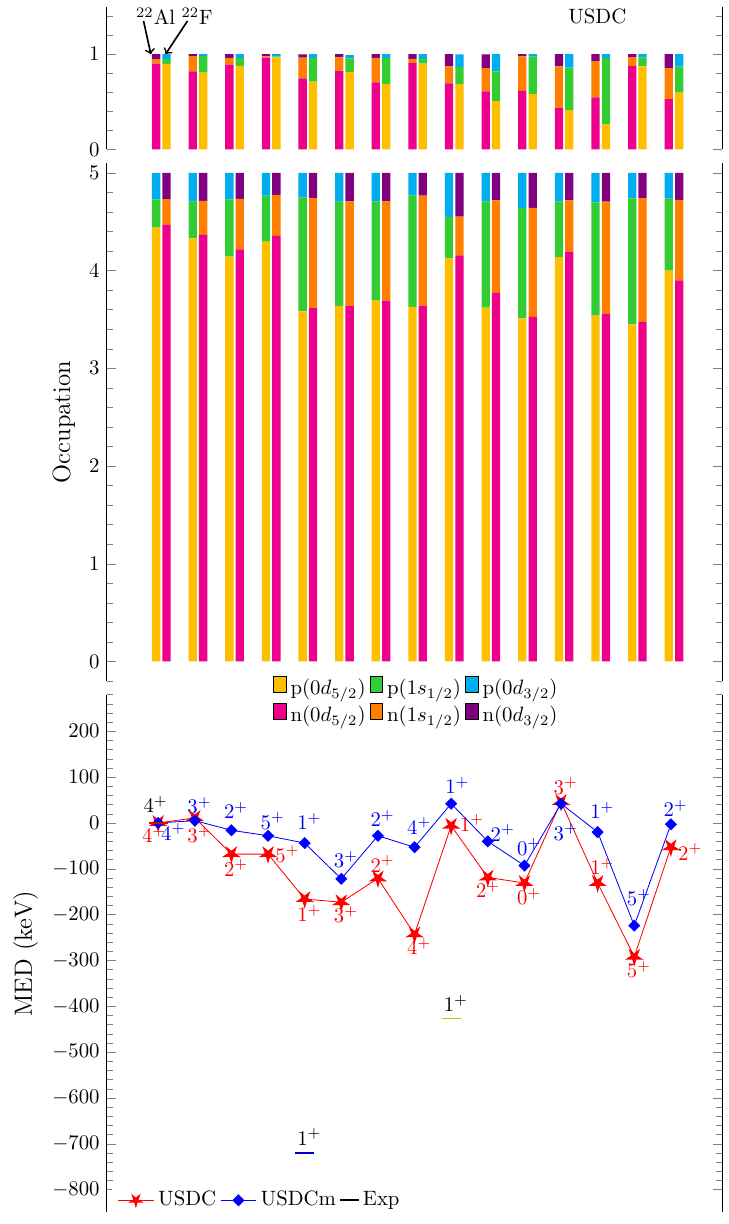}
\caption{\label{22_usdc} Comparison between the calculated and experimental \cite{NNDC} energy levels (solid lines) and the proton and neutron thresholds (dashed lines)(left), mirror energy differences for low-lying states {\color{black} and proton and neutron
occupancies of single-particle orbits for isobaric analog states of $^{22}$Al and
$^{22}$F (right).}}
\end{figure*}

{\bf $^{22}$Al/$^{22}$F:} Fig. \ref{22_usdc} shows the comparisons between the experimental data \cite{NNDC} and results obtained using USDC and USDCm interactions for the A = 22 mirror pair nuclei $^{22}$Al/$^{22}$F,
resulting in a $4^+$ ground state for $^{22}$Al and $^{22}$F, which supports the experimental assignment. The experimental levels for $^{22}$Al are taken from Ref. \cite{lee} in which three states are observed i.e., $4^+_{g.s.}$, $1^+_1$, and $1^+_2$ and in Ref. \cite{22Al_expt} $0^+_1$ state is observed which lies at a very high excitation energy, therefore we have omitted that state {\color{black} from} the spectra. Experimental data for $4^+_1$, $3^+_1$, $2^+_1$, $1^+_1$, and $1^+_2$ states in $^{22}$F are taken from Ref. \cite{22F_expt}. We observe a large deviation in the theoretical MED (calculated using the USDCm interaction) for the $1^+_1$ state of $\approx$ 670 keV in comparison to experimental MED. Although the USDC interaction minimizes this deviation in MED compared to USDCm interaction yet the deviation is greater than 550 keV in comparison to experimental MED. The $1^+_1$ state is coming from $\pi(d_{5/2}^4s_{1/2}^1)\otimes\nu(d_{5/2}^1)$ configuration using USDC interaction, presence of proton $s_{1/2}$ orbital in $^{22}$Al indicates TE shift, which leads to large MED. 
{\color{black} Although theoretically for $1^+_1$ state, we are getting a small MED which suggests that the role of continuum coupling effect is important to reproduce experimental MED in this case.}
The $1^+_2$ state is obtained from $\pi(d_{5/2}^5)\otimes\nu(d_{5/2}^1)$ configurations using USDC interaction in $^{22}$Al. {\color{black}It lacks significant proton $s_{1/2}$ occupancy, which leads to a small MED.}
Large MEDs are observed in $5^+_2$ and $4^+_2$ state  using USDC interaction, among which the average proton/neutron occupancy is {\color{black} comparatively larger} in the $s_{1/2}$ orbital for $5^+_2$ than the $4^+_2$ state therefore MED in $5^+_2$ is greater than the $4^+_2$ state. 
%Average proton/neutron occupancy in $s_{1/2}$ orbital are very high for the $5^+_2$ state in $^{22}$Al/$^{22}$F, due to which large MED is obtained for this state. The USDCm interaction also provides large MED for the $5^+_2$ state.
Due to greater proton-proton repulsion in $^{22}$Al, the single proton separation energy for $^{22}$Al is very small in comparison to the single neutron separation energy for $^{22}$F (using USDC interaction) as shown in Table \ref{sep}, which leads to unbound excited states of $^{22}$Al, whereas excited states of {\color{black}$^{22}$F are bound}. Therefore, {\color{black}$^{22}$Al is a proton-unbound system}, due to which we obtain negative MED using USDC interaction here for the most of the isobaric analog states. There is experimentally observed $B(M1)$ transition in $^{22}$F for the $2^+_1\rightarrow 3^+_1$ states and its SM calculated value by USDC interaction is 2.15 W.u., corresponding experimental value is 1.5(8) W.u. The calculated quadrupole moment for $4^+_{g.s.}$ is -0.007 $eb$ using USDC interaction, corresponding experimental value is 0.003(2) $eb$. Theoretically, we are getting a negative quadrupole moment for the ground state in $^{22}$F, whereas experimentally a positive value for the quadrupole moment is observed. The $4^+_1$ state is obtained from the $\pi (d_{5/2}^1)\otimes \nu (d_{5/2}^{-1})$ configuration i.e. one unpaired proton in $d_{5/2}$, and one neutron hole in $d_{5/2}$ orbital for $^{22}$F. If we calculate the quadrupole moment using the single particle approximation for effective charges $e_p,e_n$=(1.36,0.45)e, $Q(\text{particle})=-Q(\text{hole})$; we found that the $Q_{\text{total}}$ $(= e_{eff}^nQ(\pi d_{5/2}^{1})+e_{eff}^nQ(\nu d_{5/2}^{-1})$) is also giving a negative value for the quadrupole moment.

 In Table \ref{ft}, we have shown the calculated mirror asymmetries ($\delta$ = $|M^-_{GT}|^2 / |M^+_{GT}|^2 -1$ ) for the $1_1^+$ and $1_2^+$ states in $^{22}$F and $^{22}$Al in comparison to the earlier 
results \cite{lee}. In the case of exact isospin symmetry, mirror asymmetries $\delta$ should vanish. In this work, we have taken value of coupling constants and quenching factor as used in Ref. \cite{lee}.

\begin{table*}        
\begin{center}
%\begin{threeparttable}
\caption{Comparison of the calculated and experimental $ft$ values for the $\beta$ decay to $^{22}$F [log$ft^-$(s)] and its mirror partner $^{22}$Al [log$ft^+$(s)] using {\color{black} USDC and USDCm interactions.}}
%\label{T_3_2}
\label{ft}
%\begin{ruledtabular}
\begin{tabular}{ cccc|ccc| ccc  } 
%\hline
\hline
 & &$^{22}$O $\rightarrow $ $^{22}$F   &  & & $^{22}$Si $\rightarrow $ $^{22}$Al & & $\delta(\%)$ \\
 \hline
$J^\pi$ & Exp.  & Cal.  & Cal. \cite{lee} & Exp. & Cal.  & Cal.\cite{lee}  & Expt.   & Cal.  & Cal.\cite{lee}\\
\hline
{\color{black}
USDC} & &  &  & &  &   &    &  & \\
\hline
$1_1^+$ & 4.6(1) & {\color{black} 5.04} & 4.32  & 5.09(9) &{\color{black} 4.85} & 4.81  & 209(96) & {\color{black}-35.2}& 212\\
$1_2^+$ & 3.8(1) & {\color{black} 3.85} & 3.72  & 3.83(5) & {\color{black} 3.85} & 3.71  & 7 (28)   &{\color{black} -1.1} & -3.4\\
%\hline

\hline
USDCm & &  &  & &  &   &    &  & \\
\hline
$1_1^+$ & 4.6(1) & 5.04 & 4.32  & 5.09(9) & 4.95 & 4.81  & 209(96) & -17.4 & 212\\
$1_2^+$ & 3.8(1) & 3.86 & 3.72  & 3.83(5) & 3.84 & 3.71  & 7 (28)   & -3.3  & -3.4\\
\hline

\end{tabular}
%\end{ruledtabular}
\vspace{-10pt}
\end{center}
%\end{threeparttable}
\end{table*}

\begin{figure*}
\hspace{-1.0cm}
\includegraphics[width=8.750cm,height=10cm]{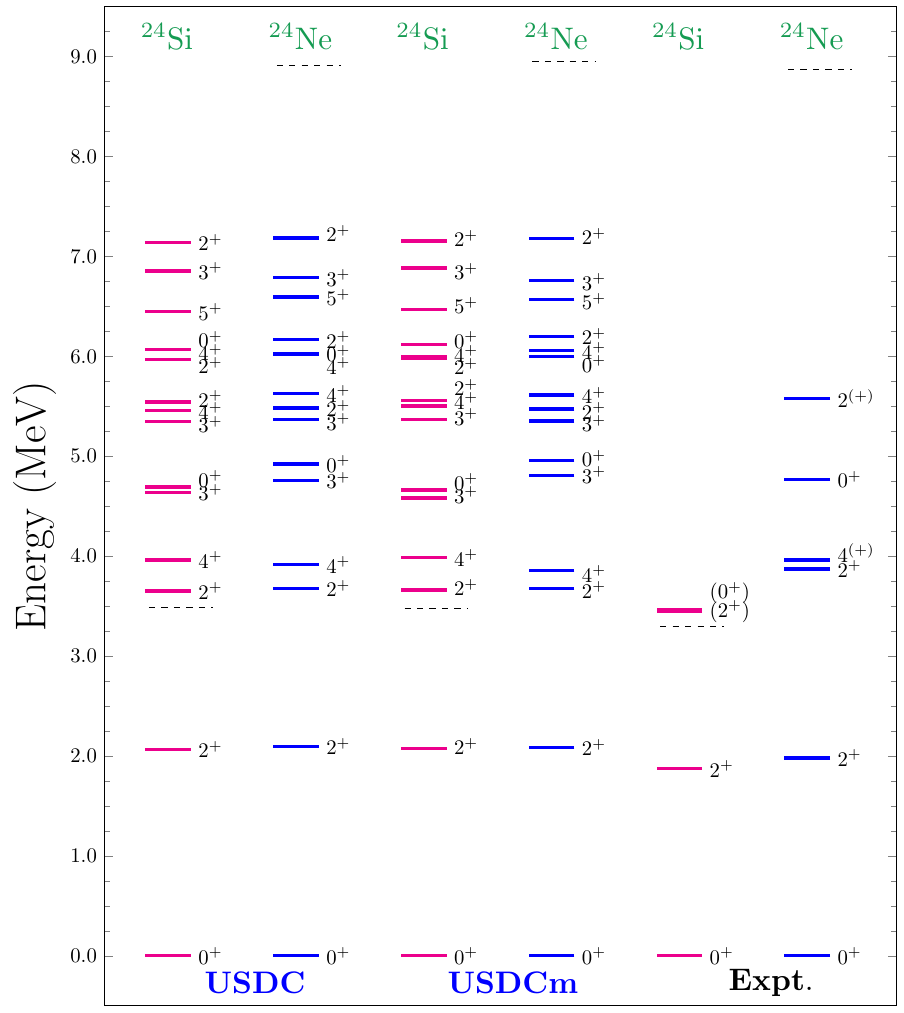}
\includegraphics[width=8.750cm,height=10cm]{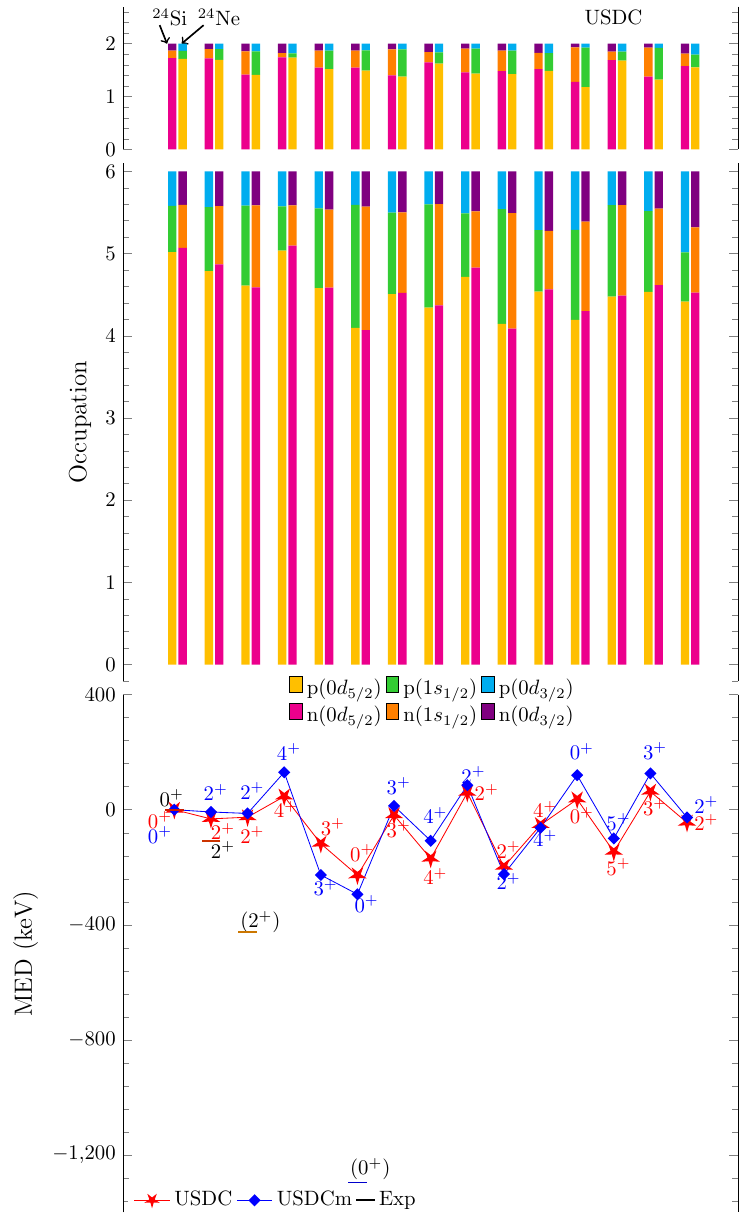}
\caption{\label{24_usdc} Comparison between the calculated and experimental \cite{NNDC} energy levels (solid lines) and the proton and neutron thresholds (dashed lines)(left), mirror energy differences for low-lying states {\color{black} and proton and neutron
occupancies of single-particle orbits for isobaric analog states of $^{24}$Si and $^{24}$Ne (right).}}
\end{figure*}

{\bf $^{24}$Si/$^{24}$Ne:} Fig. \ref{24_usdc} shows the comparisons between the experimental data \cite{NNDC} and results obtained using USDC and USDCm interactions for the A = 24 mirror pair nuclei $^{24}$Si/$^{24}$Ne,
resulting in a $0^+$ ground state for $^{24}$Si and $^{24}$Ne, which supports the experimental assignment. 
%We have calculated spectroscopic factor ($c^2S$) (using USDC interaction) for the ground state by one proton removal in $^{24}$Si, strongest transition is obtained from $d_{5/2}$ orbital, which is equal to 3.39, this indicates that ground state is solely coming from the one proton pair breaking in $d_{5/2}$ orbital, although its experimental value is $\le$ 2.8.
{\color{black} We have calculated spectroscopic factor ($C^2S$) using USDC interaction for the ground state by one proton removal in $^{24}$Si, the calculated dominant value is 3.39 for $d_{5/2}$ orbital,  corresponding experimental value is $\le$ 2.8.
 This indicates that the ground state arises solely from the one proton pair breaking in the $d_{5/2}$ orbital. }

The $0^+_1$, $2^+_1$, and $4^+_1$ state is obtained from $\pi$ ($d_{5/2})^2$/$\nu (d_{5/2})^2$ configuration in $^{24}$Si/$^{24}$Ne. Whereas $0^+_2$ state is obtained from $\pi (d_{5/2})^{-2} \otimes\nu (d_{5/2})^2$ configuration for the $^{24}$Si and $\pi (d_{5/2})^{2} \otimes\nu (d_{5/2})^{-2}$ configuration in $^{24}$Ne. The calculated B(E2;$2^+_1\rightarrow 0^+_1$) values are almost 2.5 times larger than the experimental $B(E2)$ value for the $^{24}$Si, {\color{black} this could be due to a deficiency in the calculated wavefunction.} {\color{black}Experimentally, a very large MED value equal to -1.297 MeV is observed in the tentative second $0^+$ state in $^{24}$Si and its isobaric analog state. Whereas theoretically these states are obtained at a comparatively smaller MED value of -228 and -294 keV using USDC and USDCm interactions, respectively. Which is highest after the $1/2^+_1$ state in $^{19}$Na using USDC interaction discussed in Ref. \cite{Sakshi_ISB}, which suggests a readjustment in the two-body matrix elements to reproduce this state correctly.} This state possesses a large average proton occupancy in the $s_{1/2}$ orbital, which leads to a large MED for this state. The calculated MED for the $2^+_1$ state shows good agreement with the corresponding experimental MED, as shown on the right side of Fig. \ref{24_usdc}. 
%Experimentally at 3.963 MeV excitation energy, a state with $J$=4 and tentative positive parity is observed in $^{24}$Ne, theoretically $4^+_1$ state is obtained at 3.917, and 3.837 MeV using USDC, and USDCm interactions, respectively, which are very close to experimental level, therefore SM supports the positive parity of that state.
{\color{black} For $^{24}$Ne, experimentally, the $4^{(+)}$ state is observed at 3.963 MeV excitation energy, with tentative positive parity. Theoretically, the $4^+_1$ state is obtained at 3.917 and 3.837 MeV excitation energy using USDC and USDCm interactions, respectively. Thus, the SM results suggest that this state might be the positive parity state.} 
We are able to reproduce experimental single proton ($s_{1p}$) and single neutron ($s_{1n}$) separation energies nicely from the USDC interaction. 
  
\begin{figure*}
\hspace{-1.0cm}
\includegraphics[width=8.750cm,height=10cm]{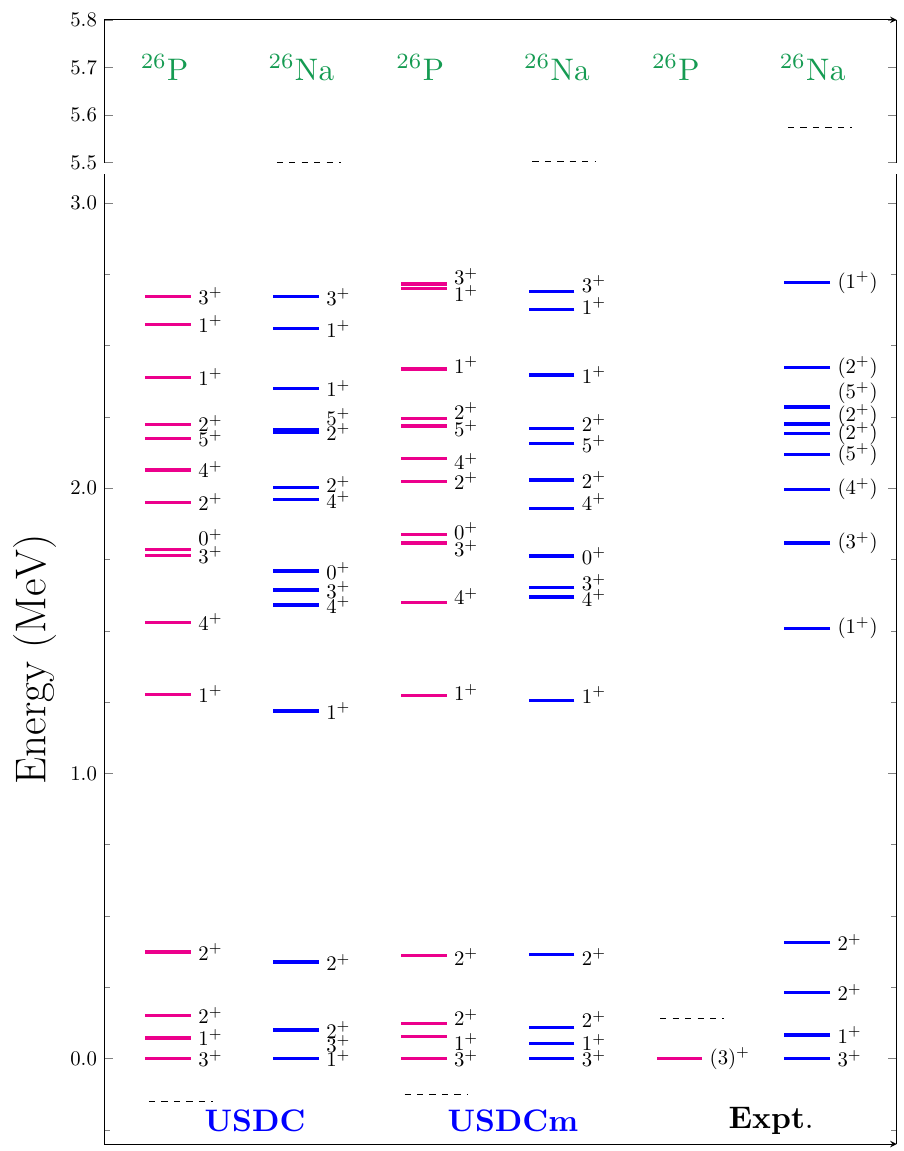}
\includegraphics[width=8.750cm,height=10cm]{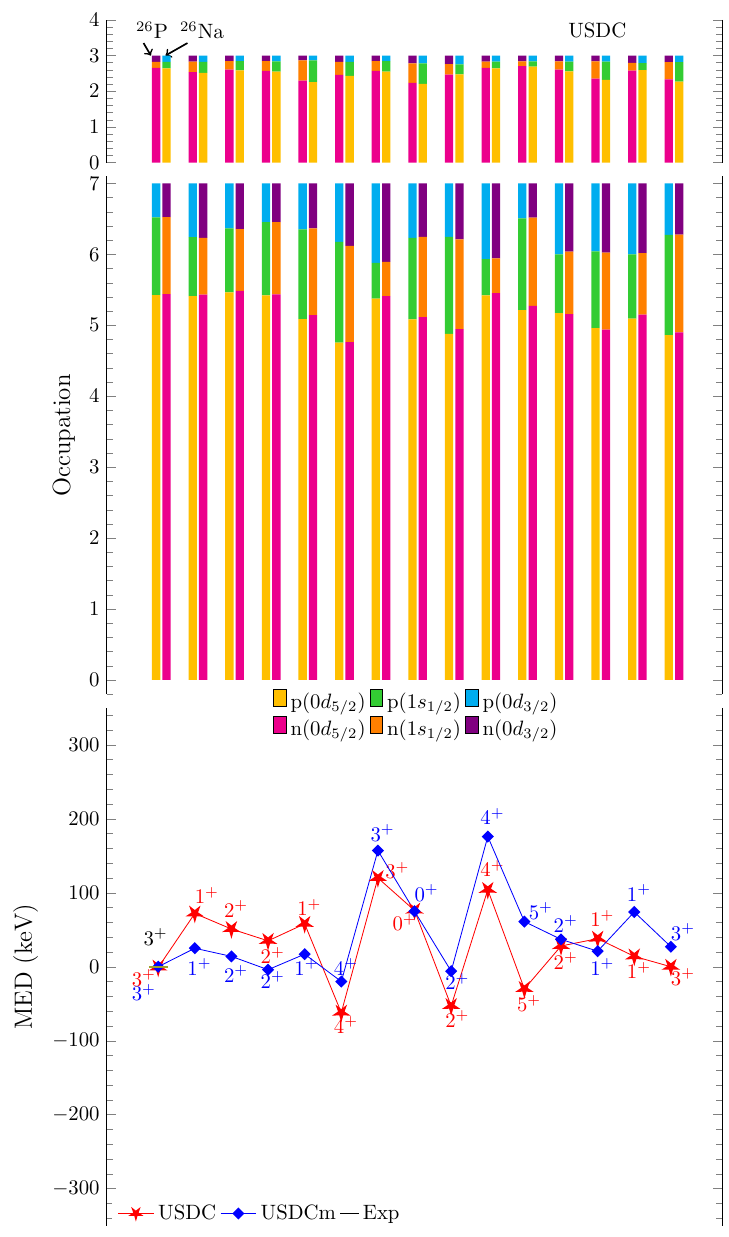}
\caption{\label{26_usdc} Comparison between the calculated and experimental \cite{NNDC} energy levels (solid lines) and the proton and neutron thresholds (dashed lines)(left), mirror energy differences for low-lying states {\color{black} and proton and neutron
occupancies of single-particle orbits for isobaric analog states of $^{26}$P and $^{26}$Na (right). }}
\end{figure*}

{\bf $^{26}$P/$^{26}$Na:} Fig. \ref{26_usdc} shows the comparisons between the experimental data \cite{NNDC} and the results obtained using USDC and USDCm interactions for the A = 26 mirror pair nuclei $^{26}$P/$^{26}$Na,
resulting in a $3^+$ ground state for $^{26}$P which supports the experimental assignment, whereas for $^{26}$Na we see inversion of the experimental ground state using the USDC interaction, resulting in $1^+$ excited state which is the experimental first excited state. Although USDCm interaction reproduces the experimental ground state correctly for $^{26}$Na. In Ref. \cite{kaneko_3} the large mirror asymmetry is evaluated for $^{26}$P/$^{26}$Na mirror pairs from the Gamow-Teller (GT) $\beta$-decay using SM by the inclusion of isospin non-conserving (INC) forces in the $sd$-shell, which shows good agreement with experimentally observed mirror asymmetry \cite{26P_halo}, which indicates $^{26}$P to be a proton halo nuclei. The calculated quadrupole moment using USDC interaction for the ground state of $^{26}$Na is -0.005 {\color{black}$eb$}, which matches very well with its experimental value i.e., -0.0053 (2) {\color{black}$eb$}. Also the calculated magnetic moment 2.64 $\mu_N$ using USDC interaction matches very well with its corresponding experimental value, which is +2.851 (2) $\mu_N$. For $^{26}$P at 0.164 MeV experimentally isomeric state with spin and parity not known is observed with a half-life 115(9) ns \cite{isomer_26P}. Assuming the transition to be pure $E2$ to the $3^+_{g.s.}$ because
an admixture of $M1$ would significantly enhance the transition rate, which suggests that the isomeric state
in $^{26}$P has spin parity either $1^+$ or $5^+$. The shell model calculations using the USDC and USDCm interactions suggest
that there are $1^+_1$
states at 0.072 and 0.077 MeV and $5^+_1$
states at 2.174 and 2.218 MeV in $^{26}$P, respectively. Since the excitation
energy of the isomeric state is comparable to that of $1^+_1$
state in the shell model calculation, therefore $1^+_1$ state is likely to be the isomeric state in $^{26}$P. Experimentally, the first $1^+$ state in $^{26}$Na is an isomeric state which decays via $E2$ transition to the yrast $3^+$ state having a half-life value 4.35 (16) $\mu s$ \cite{isomer_26P}. Theoretically calculated (from USDC interaction) half-life using $B(E2)$ value is equal to 44.73 $\mu s$.

\begin{figure*}
\hspace{-1.0cm}
\includegraphics[width=8.750cm,height=10cm]{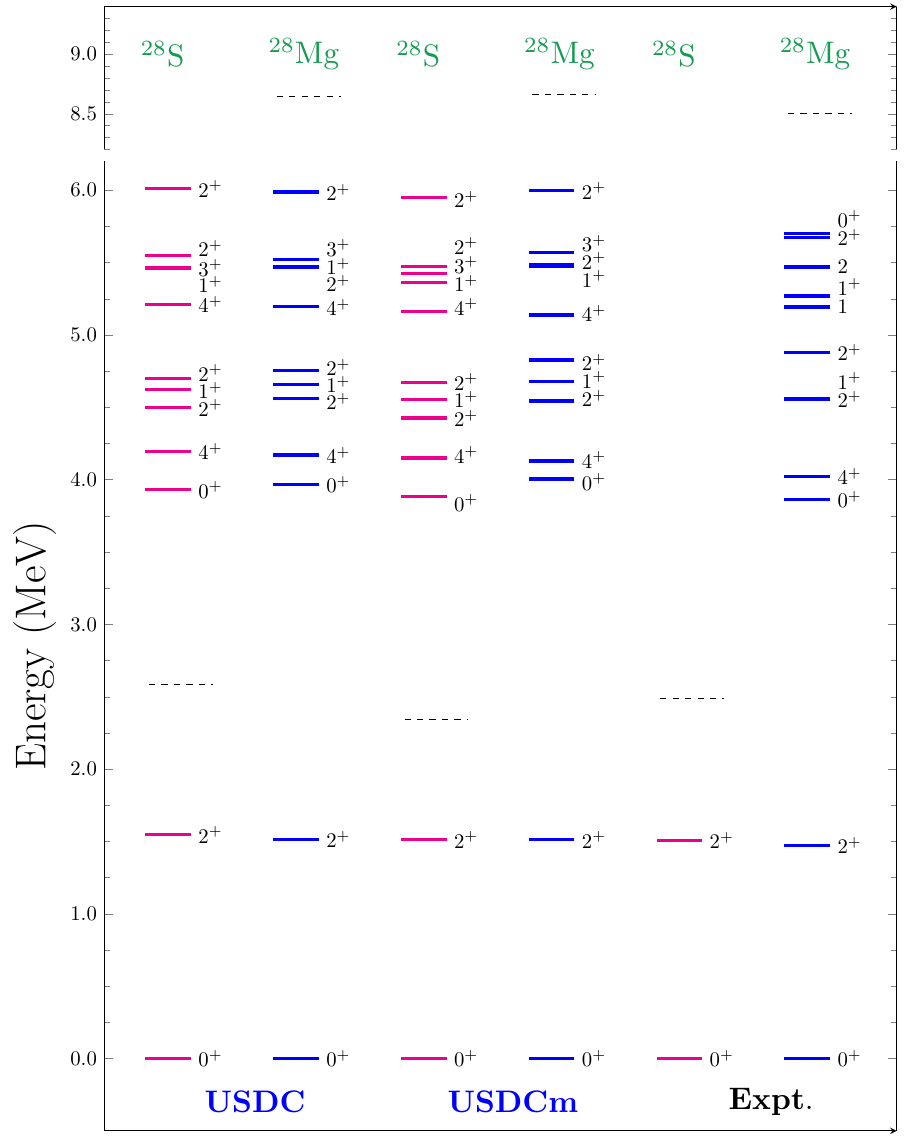}
\includegraphics[width=8.750cm,height=10cm]{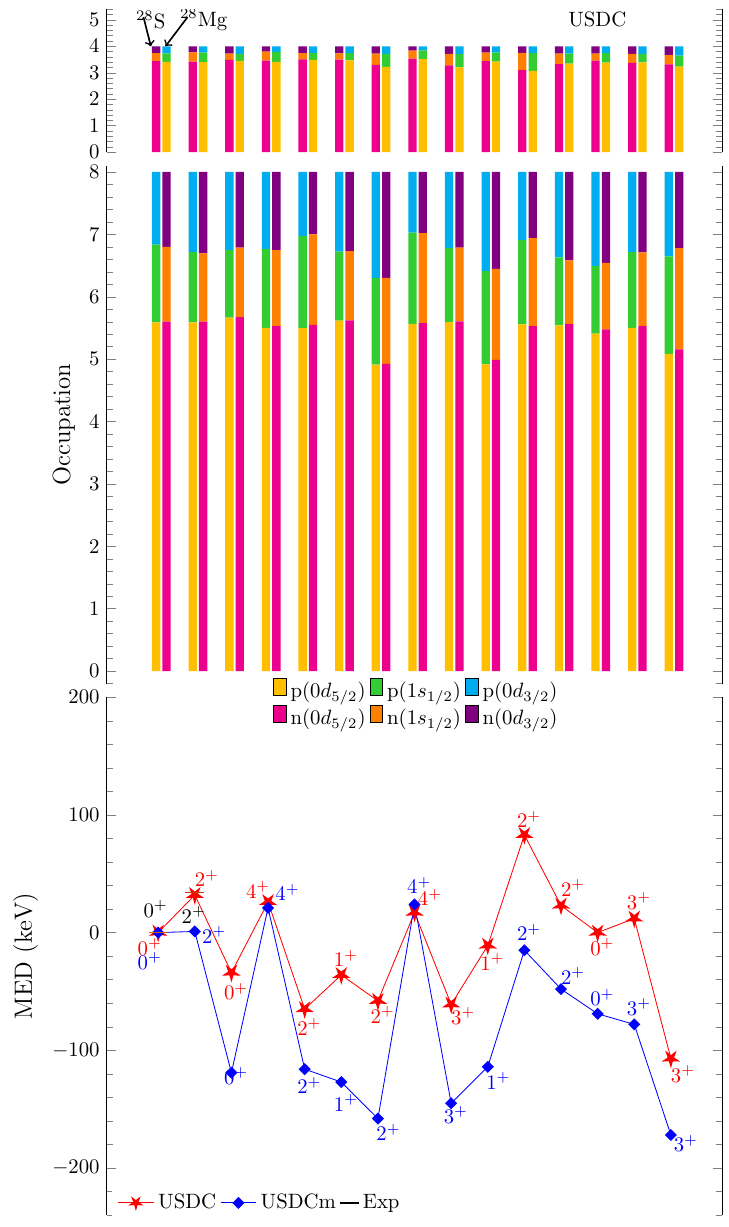}
\caption{\label{28_usdc}Comparison between the calculated and experimental \cite{NNDC} energy levels (solid lines) and the proton and neutron thresholds (dashed lines)(left), mirror energy differences for low-lying states {\color{black} and proton and neutron
occupancies of single-particle orbits for isobaric analog states of $^{28}$S and $^{28}$Mg (right). }}
\end{figure*}

{\bf $^{28}$S/$^{28}$Mg:} Fig. \ref{28_usdc} shows the comparisons between the experimental data \cite{NNDC} and the results obtained using USDC and USDCm interactions for the A = 28 mirror pair nuclei $^{28}$S/$^{28}$Mg,
resulting in a $0^+$ ground state for the $^{28}$S and $^{28}$Mg, which supports the experimental assignment. We have nicely reproduced experimental energy levels in the correct order up to 4.878 MeV excitation energy using both the interactions. For $^{28}$Mg, the ground state is obtained from the $\nu(d_{5/2}^4)$ configuration, and $1^+_1$, $2^+_1$, and $4^+_1$ states are obtained from $\pi (d_{3/2}^1s_{1/2}^1)\otimes\nu(d_{5/2}^4)$. That is, $1^+_1$ and $2^+_1$ states are formed by one proton in $d_{3/2}$ and $s_{1/2}$ orbital each, while the $4^+_1$ state comes from one unpaired proton in $d_{3/2}$ and $s_{1/2}$ orbital coupled to one neutron pair breaking in $d_{5/2}$ orbital.
% Experimentally at 5.916 MeV excitation energy tentative positive parity state $(0,1,2)^+$ is observed, when we compare this tentative level with our theoretically calculated level we found that $2^+_6$ and $0^+_4$ state lies very close, although $1^+_4$ state lies at a higher energy value.
On the left side of Fig. \ref{28_usdc} we can see that both of the interactions are giving large MED for $3^+_3$ state, due to the large average proton/neutron occupancy in $s_{1/2}$ orbital. Experimentally $E2$ transition is observed between $0^+_2\rightarrow2^+_1$ and $4^+_1\rightarrow2^+_1$ in $^{28}$Mg with the observed value 13.1 (20) and 50 (20) $e^2fm^4$, corresponding calculated B(E2;$0_2^+\rightarrow2^+_1$) and B(E2;$4_1^+\rightarrow2^+_1$) values are 6.2 and 70.80 $e^2fm^4$ using USDC interaction and 6.10 and 68.54 $e^2fm^4$ using USDCm interaction, respectively. Experimental mixing ratio for the $M1$+$E2$ transition between $2^+_2\rightarrow2^+_1$ and $2^+_3\rightarrow2^+_1$ are +0.04 (3) and +0.35 (6), respectively. The corresponding calculated values are 0.0021 and -0.51 using USDC interaction and -0.002 and -0.64 using USDCm interaction, respectively. 

\begin{figure*}
\hspace{-1.0cm}
\includegraphics[width=8.750cm,height=10cm]{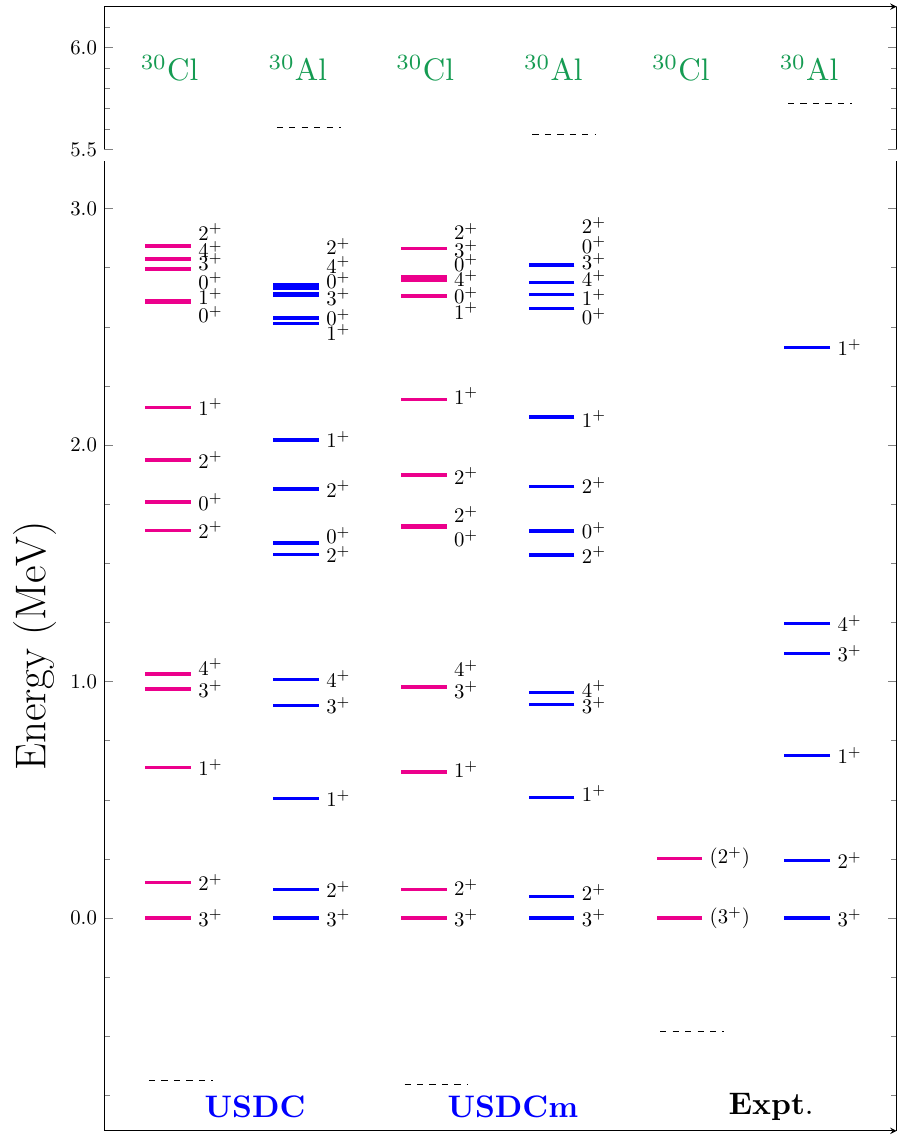}
\includegraphics[width=8.750cm,height=10cm]{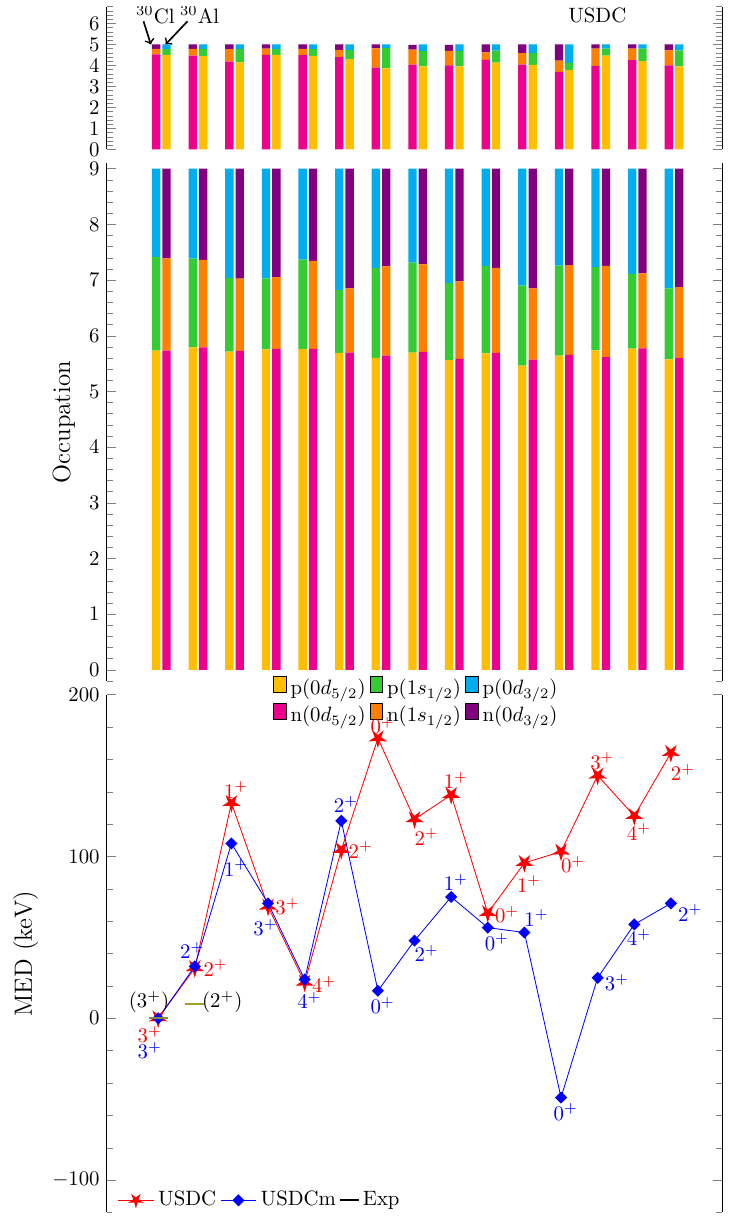}
\caption{\label{30_usdc} Comparison between the calculated and experimental \cite{NNDC} energy levels (solid lines) and the proton and neutron thresholds (dashed lines)(left), mirror energy differences for low-lying states {\color{black} and proton and neutron
occupancies of single-particle orbits for isobaric analog states of $^{30}$Cl and $^{30}$Al (right). }}
\end{figure*}

{\bf $^{30}$Cl/$^{30}$Al:} Fig. \ref{30_usdc} shows the comparisons between the experimental data \cite{NNDC} and results obtained using USDC and USDCm interactions for the A = 30 mirror pair nuclei $^{30}$Cl/$^{30}$Al,
resulting in a $3^+$ ground state for $^{30}$Cl and $^{30}$Al, which supports the experimental assignment. Experimental data for the $^{30}$Al are taken from Ref. \cite{30Al_expt}. There is only a ground state, and the first experimental state observed for the $^{30}$Cl although both are tentative. For $^{30}$Al, we have successfully reproduced experimental states up to 0.7 MeV excitation energy by SM. The $3^+_{g.s.}$, $2^+_1$, and $4^+_1$ states in $^{30}$Cl are obtained from $\pi(d_{3/2}^1)\otimes\nu(d_{5/2}^5)$. The $0^+_1$ and $1^+_1$ states are obtained from the $\pi (d_{3/2}^1)\otimes\nu (d_{5/2}^4s_{1/2}^1$) configurations, whereas $2^+_1$ state is obtained from $\pi (d_{3/2}^1)\otimes\nu (d_{5/2}^5$) configuration. We have reproduced the MED for $2^+_1$ mirror state in reasonable agreement with the experimental value. The SM gives the highest MED for the $0^+_1$ state of 173 keV using USDC interaction and USDCm interaction gives the highest MED for $2^+_2$ state of 122 keV. Experimentally, the $M1$ cascade is observed in $^{30}$Al for $1^+\rightarrow2^+\rightarrow3^+$ with experimental value 0.62$_{+25}^{-14}$ and 0.18$_{+6}^{-4}$ $\mu_N^2$ corresponding theoretical values are 0.62 and 0.10 $\mu_N^2$, respectively. Experimentally for the $^{30}$Al, $1^+_1$ state decays to $2^+_1$ and $3^+_{g.s.}$ states with the branching ratio of 98 and 2$\%$ \cite{30Al_expt}, respectively. Theoretically, by assuming $M1$ and $E2$ transition between $1^+_1\rightarrow2^+_1$ and $E2$ transition between $1^+_1\rightarrow3^+_1$, the calculated branching ratios are 99.86 and 0.14$\%$ using the USDC interaction, respectively. Experimentally, the $3^+_2$ state in $^{30}$Al decays to $2^+_1$ and $3^+_{g.s.}$ states with branching ratios 69 and 31$\%$, respectively. Theoretically calculated (using USDC interaction) branching ratios for $3^+_2\rightarrow2^+_1$ and $3^+_2\rightarrow3^+_{g.s.}$ transitions from $B(M1)$ and $B(E2)$ values are 76.15 and 23.85$\%$, respectively. 

\begin{figure*}
\hspace{-1.0cm}
\includegraphics[width=8.750cm,height=10cm]{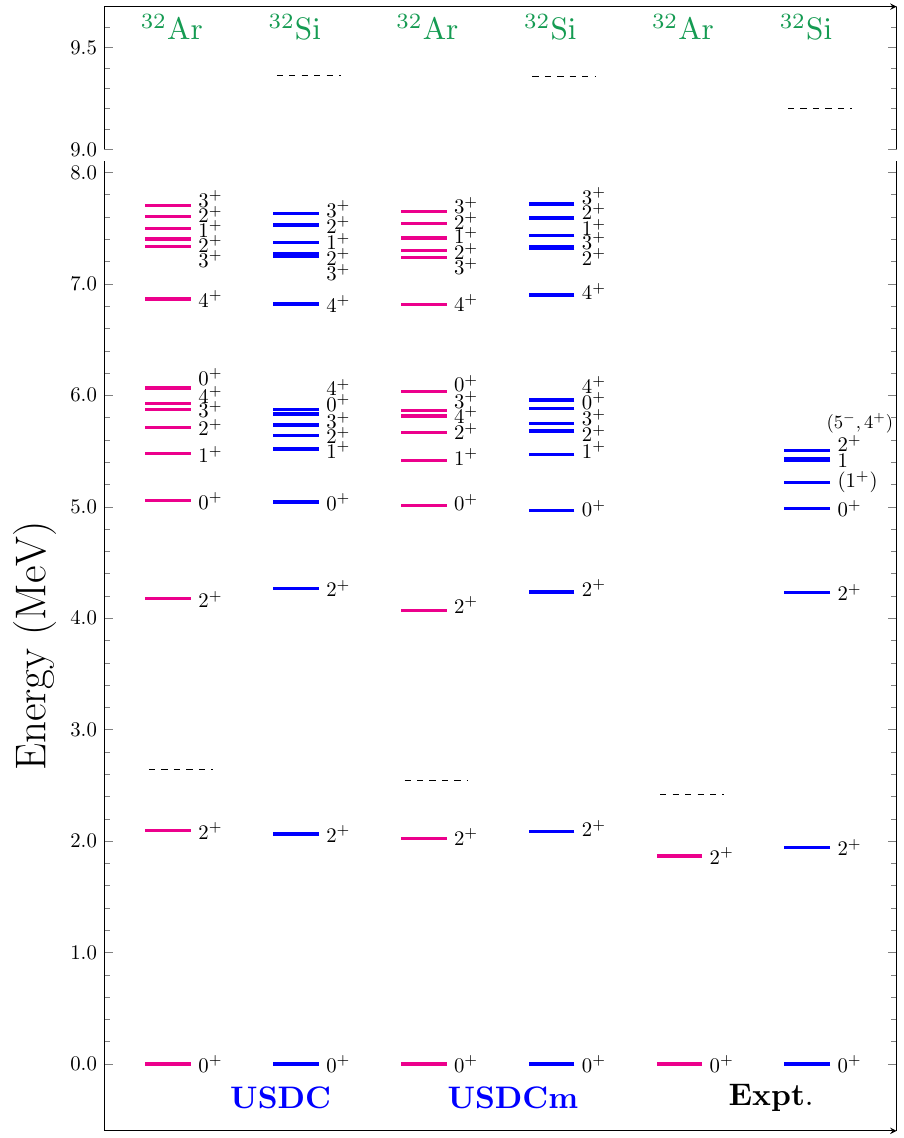}
\includegraphics[width=8.750cm,height=10cm]{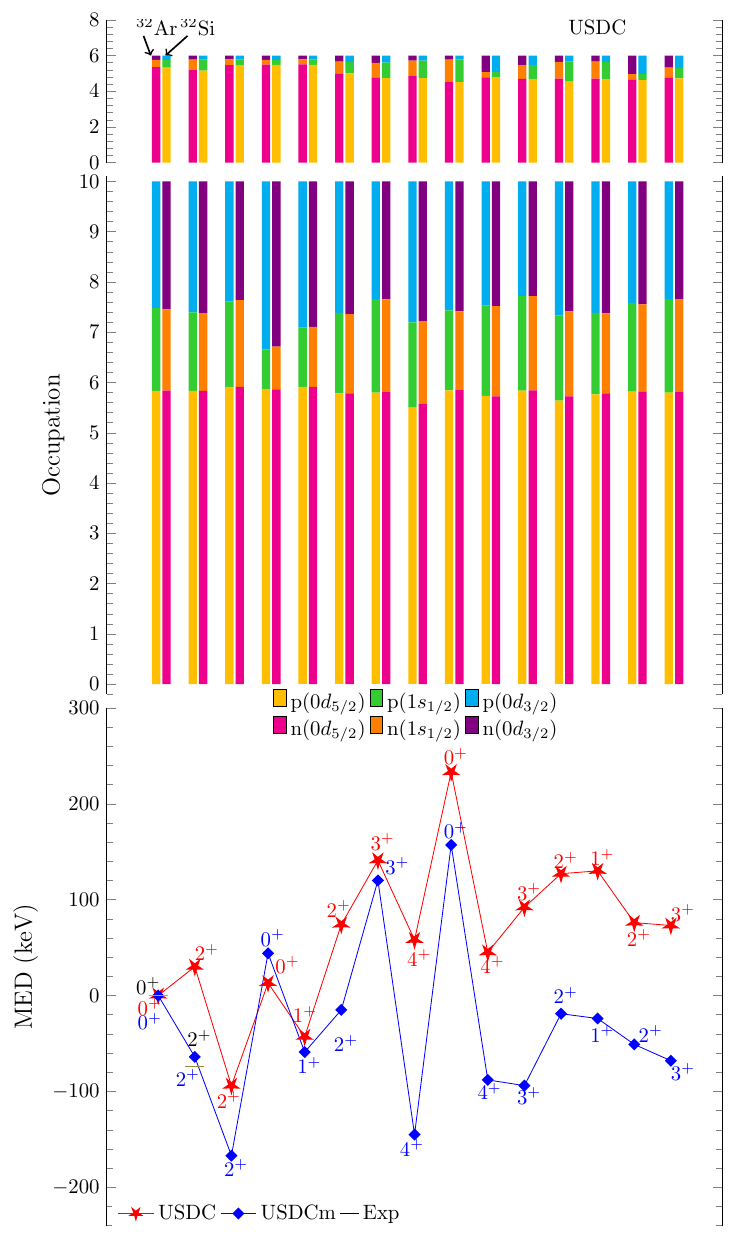}
\caption{\label{32_usdc} Comparison between the calculated and experimental \cite{NNDC} energy levels (solid lines) and the proton and neutron thresholds (dashed lines)(left), mirror energy differences for low-lying states {\color{black} and proton and neutron
occupancies of single-particle orbits for isobaric analog states of $^{32}$Ar and $^{32}$Si (right).}}
\end{figure*}

{\bf $^{32}$Ar/$^{32}$Si:} Fig. \ref{32_usdc} shows the comparisons between the experimental data \cite{NNDC} and the results obtained using the USDC and USDCm interactions for the A = 32 mirror pair nuclei $^{32}$Ar/$^{32}$Si,
resulting in a $0^+$ ground state for $^{32}$Ar and $^{32}$Si, which supports the experimental assignment. There are {\color{black} few} experimental data available for $^{32}$Ar, we have shown only two confirmed states here. Tentative experimental states $(0^+,2^+)$ at very high excitation energies are available for $^{32}$Ar, we have omitted those states here. We have successfully reproduced all the experimental states up to 4.984 MeV excitation energy for $^{32}$Si. Experimentally, at 5.219 MeV excitation energy a tentative state $(1^+)$ is available in $^{32}$Si. By comparing this energy level with the theoretically calculated energy levels for $1^+_1$ state at 5.518 and 5.471 MeV energies using USDC and USDCm interactions, respectively. We can predict this tentative state to be $1^+_1$. The ground state and first excited state in $^{32}$Ar is obtained from $\pi (d_{3/2}^2)$ configuration, i.e., one proton pair breaking in the $d_{3/2}$ orbital. The yrast $3^+$ and $4^+$ states are obtained from the $\pi (d_{3/2}^2)\otimes\nu (d_{5/2}^5s_{1/2}^1$) and $\pi (d_{3/2}^3d_{5/2}^5)$ configurations, respectively. We observe a large positive MED for $0^+_3$ state using both USDC and USDCm interactions. Experimentally, the quadrupole moment for the first $2^+$ state is observed equal to 0.11 (10) eb for the $^{32}$Si, corresponding SM result using USDC and USDCm interactions is 0.13 eb, which shows very good agreement with the experimental value. There is $M1$+$E2$ transition observed experimentally for $2^+_2\rightarrow2^+_1$ in $^{32}$Si. We have calculated the mixing ratio for this transition using SM, {\color{black} which is equal to -0.46 and -0.37 from USDC and USDCm interactions, respectively. The corresponding experimental value is -0.84 (44).} 
%There is E2 cascade observed experimentaly between $$

\begin{figure*}
\hspace{-1.0cm}
\includegraphics[width=8.750cm,height=10cm]{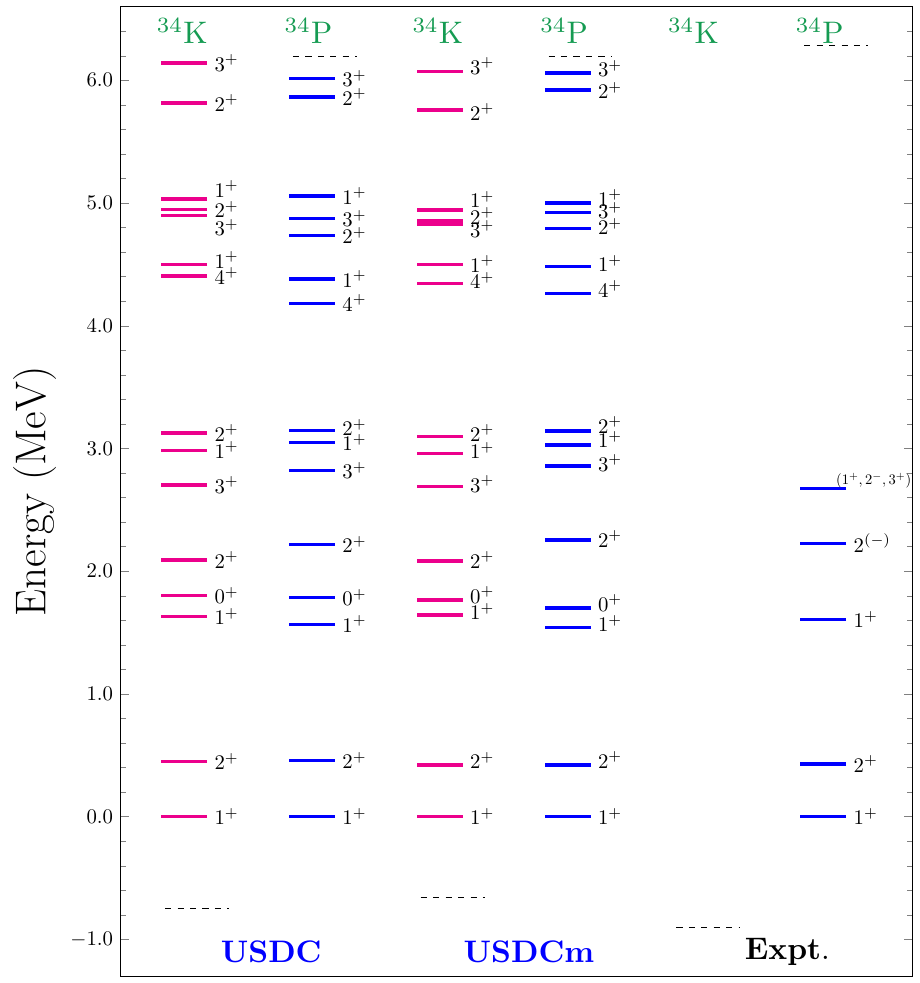}
\includegraphics[width=8.750cm,height=10cm]{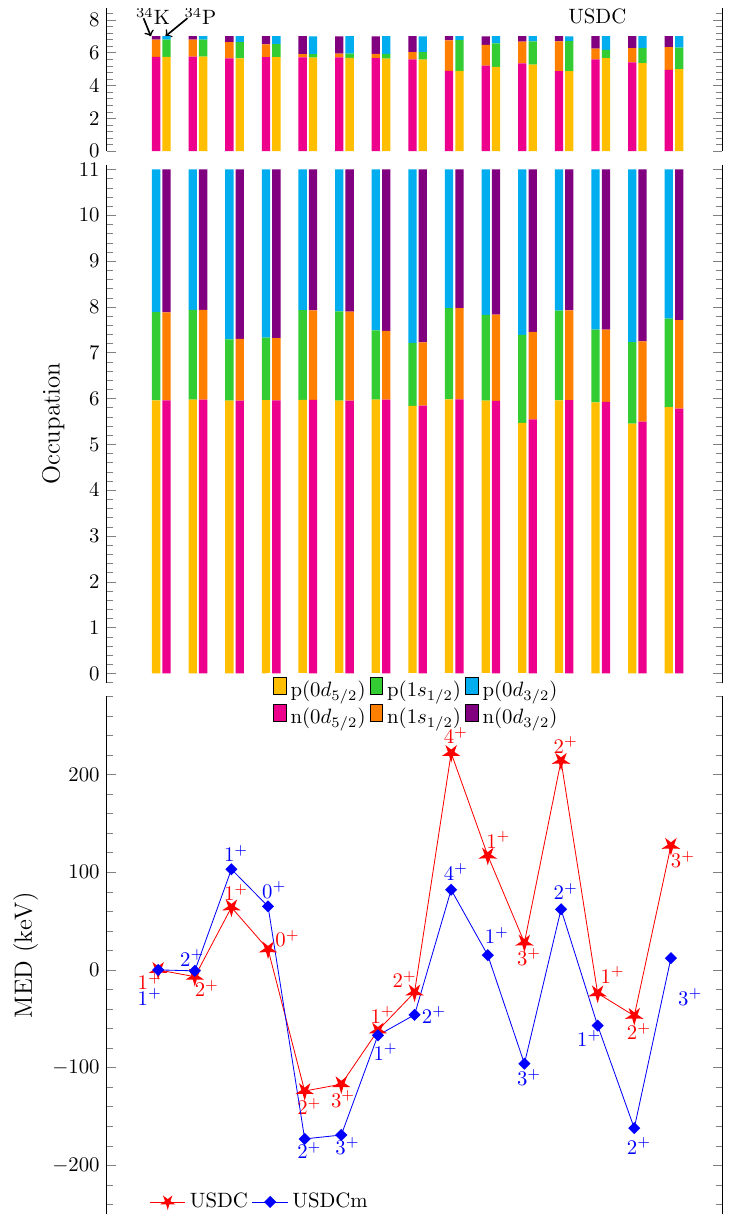}
\caption{\label{34_usdc} Comparison between the calculated and experimental \cite{NNDC} energy levels (solid lines) and the proton and neutron thresholds (dashed lines)(left), mirror energy differences for low-lying states {\color{black} and proton and neutron
occupancies of single-particle orbits for isobaric analog states of $^{34}$K and $^{34}$P (right).}}
\end{figure*}

{\bf $^{34}$K/$^{34}$P:} Fig. \ref{34_usdc} shows the comparisons between the experimental data \cite{NNDC} and the results obtained using USDC and USDCm interactions for the A = 34 mirror pair nuclei $^{34}$K/$^{34}$P,
resulting in a $1^+$ ground state for $^{34}$P, which supports the experimental assignment. Whereas no experimental data are available for the $^{34}$K. The SM reproduces the first three experimental levels very well for $^{34}$P. 
The ground state and first excited state of $^{34}$P is formed by $\pi (s_{1/2}^1)\otimes \nu (d_{3/2}^{-1})$ doublet i.e., both of the states share almost similar configuration. The second and first $0^+$ states are $\pi (s_{1/2}^1)\otimes \nu (s_{1/2}^{1})$ doublet. The $3^+_1$ state is obtained from the $\pi (d_{3/2}^1)\otimes \nu(d_{3/2}^3)$ configuration in $^{34}$P. The experimental mixing ratio for the $2^+\rightarrow 1^+$ transition in $^{34}$P is +0.11$_{-12}^{+13}$, the corresponding calculated values for $2^+_1\rightarrow1^+_1$ are -0.003 and -0.002 using USDC and USDCm interactions, respectively. Experimentally there is $M1$+$E2$ cascade in $^{34}$P $1^+_2\rightarrow2^+_1\rightarrow1^+_1$ and $1^+_2\rightarrow1^+_1$ \cite{34P_expt} with experimental $B(M1)$ values 0.014$_{-14}^{10}$, 0.37$_{-18}^{+9}$ and 0.0064$_{-55}^{18}$ $\mu_N^2$, respectively. The experimental $B(E2)$ values for the same transitions are 157$_{-157}^{105}$, 327$_{-327}^{850}$ and 0.6 (6) $e^2fm^4$, respectively. From the USDC interaction, the calculated $B(M1)$ values for the same transitions are 0.0067, 0.580, and 0.0036 $\mu_N^2$, and the $B(E2)$ values are 26.35, 0.40, and 4.70 $e^2fm^4$, respectively. The mixing ratios for the $2^+_1\rightarrow1_{g.s.}^+$, $1^+_2\rightarrow2^+_1$, and $1^+_2\rightarrow1^+_{g.s.}$ transitions in $^{34}$P using USDC interaction are -0.003, 0.58, and 0.47, respectively. The corresponding experimental values are +0.11$_{-12}^{+13}$, +1.0 (6) and -0.13 (6), respectively.

\begin{figure*}
\hspace{-1.0cm}
\includegraphics[width=8.750cm,height=10cm]{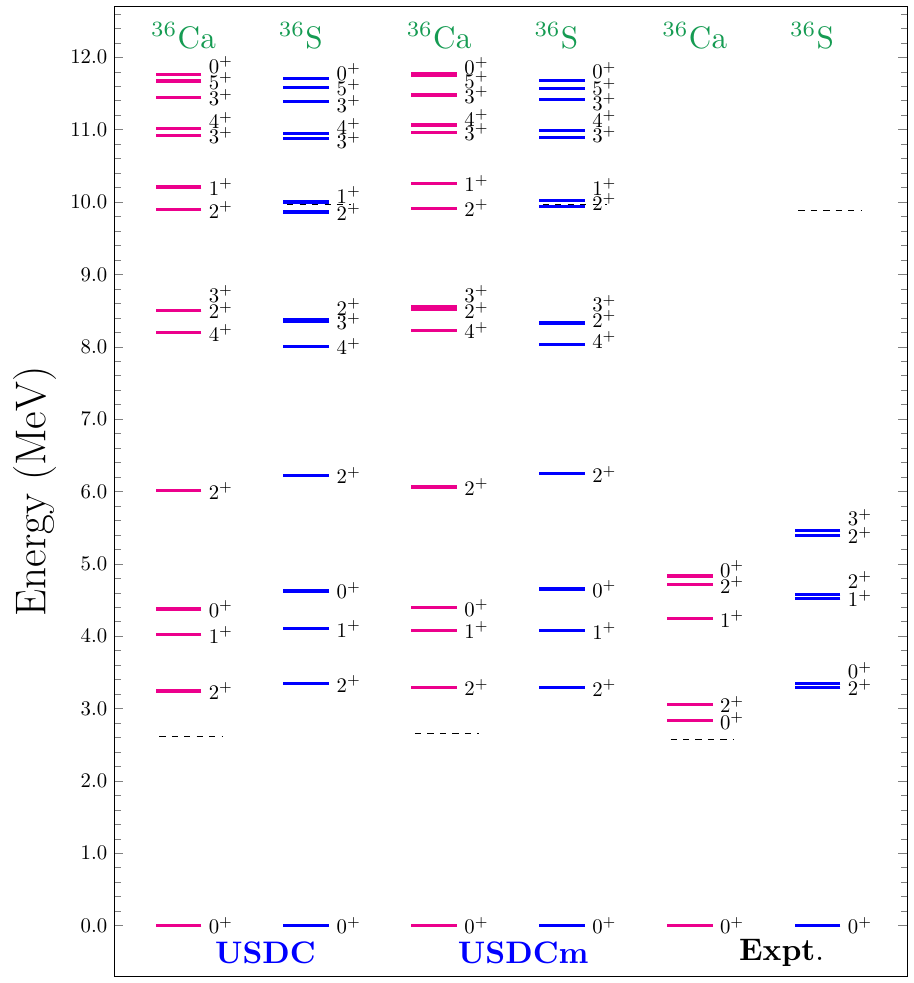}
\includegraphics[width=8.750cm,height=10cm]{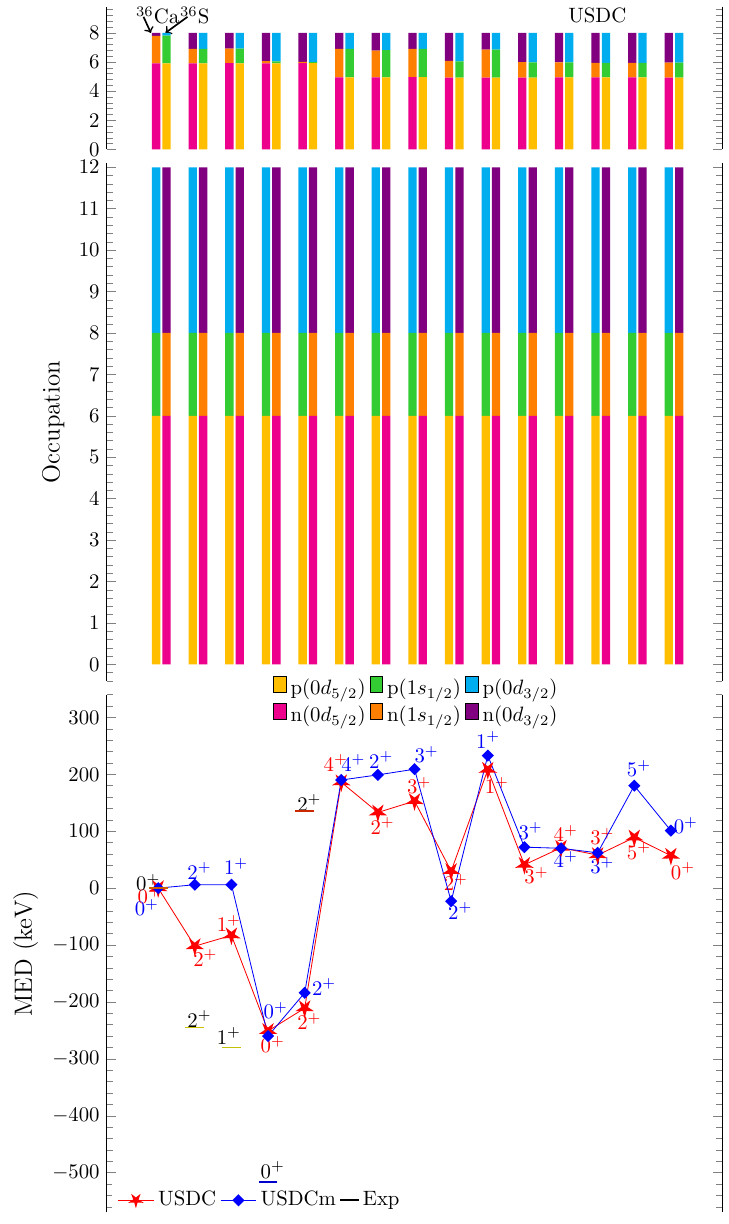}
\caption{\label{36_usdc} Comparison between the calculated and experimental \cite{NNDC} energy levels (solid lines) and the proton and neutron thresholds (dashed lines)(left), mirror energy differences for low-lying states {\color{black} and proton and neutron
occupancies of single-particle orbits for isobaric analog states of $^{36}$Ca and $^{36}$S (right).}}
\end{figure*}

{\bf $^{36}$Ca/$^{36}$S:} Fig. \ref{36_usdc} shows the comparisons between the experimental data \cite{NNDC} and the results obtained using the USDC and USDCm interactions for the A = 36 mirror-pair nuclei,
resulting in a $0^+$ ground state for the $^{36}$Ca and $^{36}$S, which supports the experimental assignment. Experimentally, the third $2^+$ and first $3^+$ states are obtained to be very close, whereas these states are lying at a very high gap in excitation energy. In the Ref. \cite{36Ca_expt} intruder state $0_2^+$ is observed at 2.83 (13) MeV excitation energy, which appears below the first $2^+$ state, although we are not able to reproduce this intruder state. This intruder state is reproduced below $2^+_1$ state by SM calculations using the $sd-pf$ model space in Ref. \cite{36Ca_expt} by exciting two protons (neutrons) across the $Z=20$ shell closure for $^{36}$Ca ($^{36}$S). The third $0^+$ state in $^{36}$Ca is observed experimentally at 4.83 (17) MeV, which is obtained at 11.760 and 11.780 MeV, using USDC and USDCm interactions, respectively. This state is obtained from the $\nu(d_{5/2}^4s_{1/2}^2)$ configuration. In this case large MED is obtained for $0^+_2$, and $2^+_2$ isobaric analog states using both the interactions. The experimental energy levels for $^{36}$S is taken from the Ref. \cite{36S_expt}. We obtain a large deviation in the MED value for $2^+_2$ state corresponding to its experimental value almost equal to 346 keV, whereas this deviation was obtained almost 860 keV from the SM using the $sd-pf$ model space \cite{36Ca_expt}. Results obtained for MED using USDC interaction for the $2^+_1$ and $1^+_1$ states are closer to experimental MED in comparison to the USDCm. The $1^+_1$ and $2^+_1$ states are coming from the $\nu(d_{3/2}^1s_{1/2}^{-1})$ configuration in $^{36}$Ca and $\pi(d_{3/2}^1s_{1/2}^{-1})$ configuration in $^{36}$S isotopes. Due to a pure 1p-1h configuration, the observed shifts in energy are almost identical in this mirror pair for these states.

\begin{figure}
\includegraphics[width=0.47\linewidth]{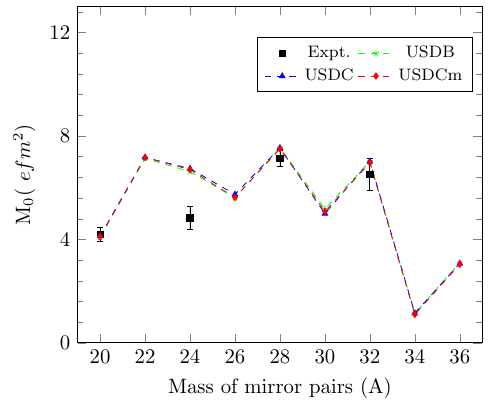}
\includegraphics[width=0.4\linewidth]{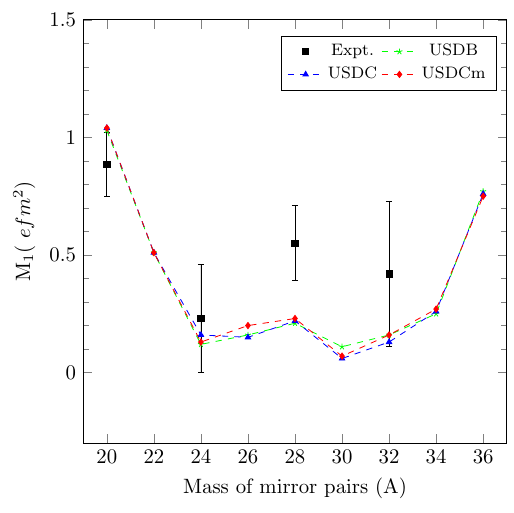}
\caption{Comparison of the experimental \cite{NNDC,Togano,32Ar_expt,be2_36Ca} and theoretical isoscalar ($M_0$) (top) and isovector ($M_1$) matrix element (bottom) for $T_z=\pm2$ mirror pair.}
\label{fig:iso}
\end{figure}

\begin{figure}
\centerline{\includegraphics[width=0.5\linewidth]{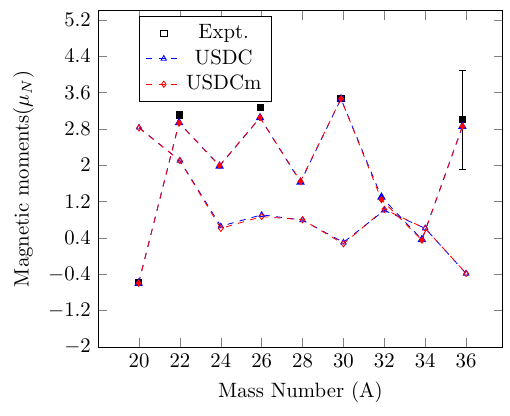}}
\caption{Expt. \cite{NNDC}, USDC, and USDCm results for magnetic moment is shown by unfilled (filled) for $T_{z}=-2$ ($T_{z}=+2$) square, triangle and diamond symbols, respectively.}
\label{mag_moment}
\end{figure}

\begin{figure*}
\centering
\includegraphics[width=0.4\linewidth]{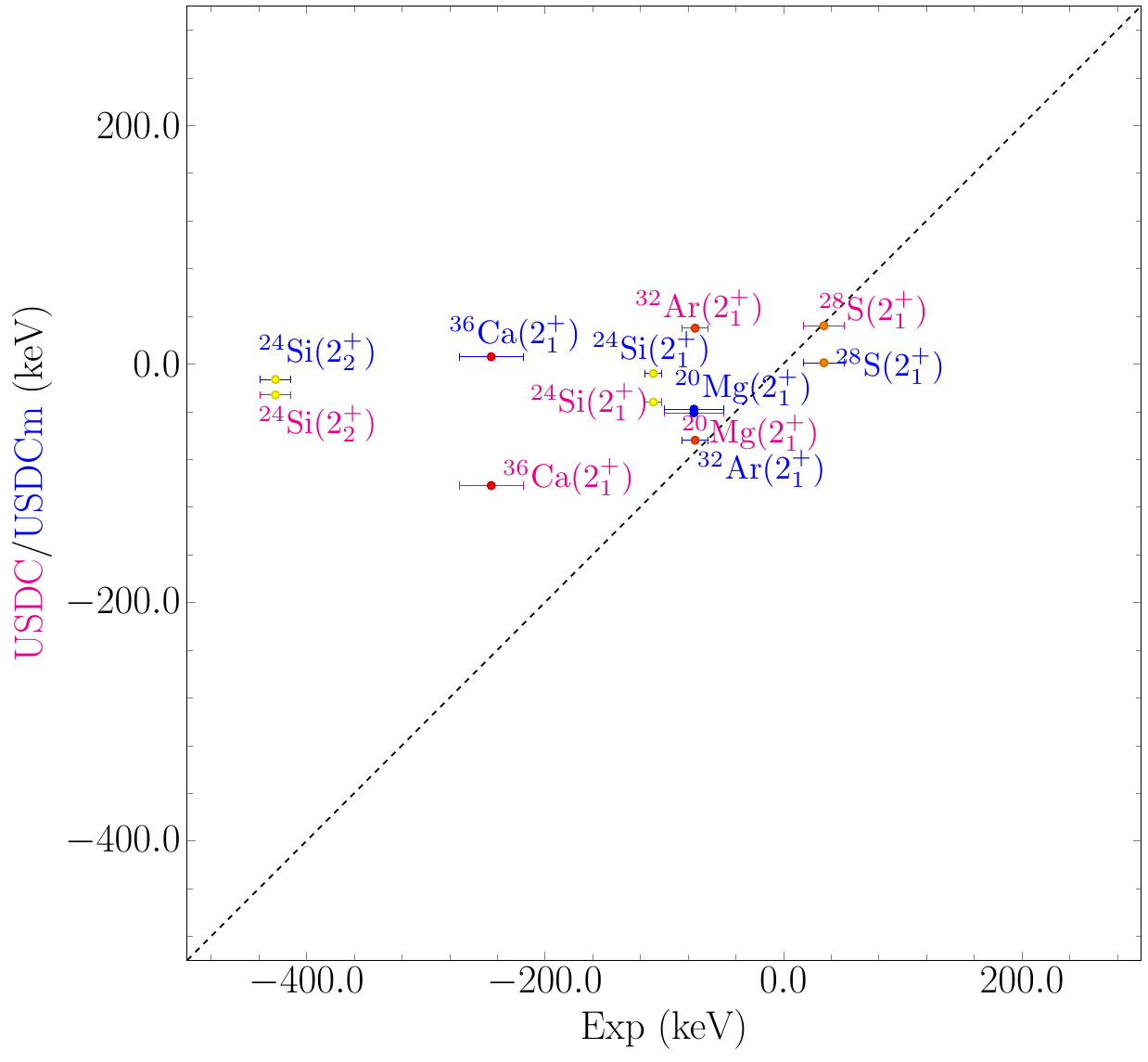}
\caption{\label{sd}The calculated MEDs with USDC (in red color) and USDCm (in blue color) interactions, labeled with the name of the associated proton-rich nucleus, are
compared with the experimental data with error bars.}
\end{figure*}

In Table \ref{T_3_2}, we can see that electromagnetic transition strength $B(E2)$ is available for all even-even $T_{z} = 2$ nuclei at or near the valley of stability. The results from the VS-IMSRG interaction reported in Table \ref{T_3_2} is taken from the Ref. \cite{E2_sd}. The theoretically calculated $B(E2)$ values show better agreement with the experimental data in comparison to the VS-IMSRG interaction (data taken from \cite{E2_sd}). The calculated MEDs for the $A=20$ and $32$ ($T_z=\pm 2$) mirror pairs, $^{28}$S ($T_z=+2$) and $^{36}$S ($Tz=+2$) nuclei show very good agreement with the experimental values. The calculated $B(E2)$ values are very close to those obtained from the {\color{black}isospin non-conserving} and isospin conserving $sd$- shell interactions: USDB, USDC, and USDCm, which show $B(E2)$ values are not a very sensitive quantity towards ISB effects. The $B(E2)$ transition mainly depends on the collective quadrupole motion of the nucleons. For the $sd$-shell nuclei isoscalar ($M_0$) and isovector ($M_1$) matrix element corresponding to $E2$ transition provides a more refined test of the theory for the $B(E2)$ values. Therefore, we have calculated $M_0$ and $M_1$ using USDB, USDC, and USDCm interactions and compared them with their available experimental data, as shown in Fig. \ref{fig:iso}. Fig. \ref{fig:iso} shows that the results obtained from the USDB interaction do not show much deviation from the results obtained from the USDC and USDCm interactions. This indicates that the isovector matrix element obtained from the isospin-conserving interaction is the same as that of isospin {\color{black} non-conserving} interaction, which shows that ISB does not play a significant role in $B(E2)$ values. The calculated isoscalar ($M_0$) matrix elements show good agreement with experimentally available data for the $A=20$, $28$, and $32$ mirror pairs. The calculated isovector ($M_1$) matrix elements show good agreement with experimentally available data for the $A=20$, $24$, and $32$ mirror pairs. For the $A$=24, the theoretical $M_1$ value matches well with the experimental $M_1$ value, whereas the theoretical $M_0$ value is overestimated by a factor $\approx$1.4. While for the $A=28$, the theoretical $M_0$ result shows good agreement with the experimental $M_0$ value, and the theoretical $M_1$ value is underestimated by a factor $\approx$ 2.5. \\ The theoretical magnetic moment corresponding to the USDC and USDCm interactions has been plotted in Fig. \ref{mag_moment} in comparison with their available experimental data. For the odd-odd mirror nuclei, we have plotted the magnetic moment of the ground state, whereas for even-even mirror nuclei we have plotted the magnetic moment for the first excited state. For $^{26}$Na, experimentally there is a tentative ground state $(3)^+$, although the USDC interaction provides $1^+_{1}$ as a ground state, therefore for comparison with the experimental magnetic moment, we have taken $3^+_1$ using the USDC interaction instead of $1^+_{g.s}$. The calculated result shows good agreement with the available experimental data.

 The TE shift \cite{TE_1,TE_2} that can be observed in mirror pairs due to the Coulomb energy differences in mirror states is one of the most crucial features for understanding the asymptotic behavior of mirror states. The mirror states of those $sd$-shell nuclei, which are near the proton dripline, exhibit significant TE shifts, discussed in these Ref. \cite{zhang,lee,19Na, MED_22Al}. When we consider the $0^+_2$ isobaric analog state in $A$=24, $T_z=\pm2$ mirror pair, we find that the deviation in theoretical value (using USDC interaction) with experimental value is 1069 keV, which is the highest among all the $T_{z}$=$\pm 2$, $sd$-shell nuclei. For the $1^+_1$ isobaric analog state of $A=22$ mirror pair, a large discrepancy is observed in its theoretically calculated (using USDC interaction) and experimentally observed MED value. Whereas USDCm interaction minimizes the deviation from experimental MED. If continuum coupling effect is included in Gamow shell model (GSM) with EM1.8/2.0 interaction, then the obtained difference between experimental and theoretical MED is not very large as discussed in Ref. \cite{MED_22Al}. Also, all the states of $^{22}$Al obtained from GSM calculations show resonances. The experiment \cite{lee} detected proton emissions from the two $1^+$ states in $^{22}$Al, indicating the resonances of the $1^+$ states. Incorporating isospin non-conserving interactions into the USD shell-model interaction should provide a more accurate description of nuclear structure, particularly by mimicking the effects of continuum coupling. This is especially relevant for configurations involving the $s_{1/2}$ orbital, which strongly couples to the continuum due to its weak binding and low angular momentum. In contrast, standard SM calculations using the USDC interaction tend to underestimate mirror energy differences, which shows a significant role of continuum coupling and INC effects in reproducing experimental observations.

{\color{black} It is important to mention here that the predictive power for low-lying states MEDs in some cases is not as strong as in the experimental data.  This may be due to the fitting including less data from excited states.}
%If we include a new data set for more excited states in the fitting, then the prediction of MEDs might be better.}

\section{Summary and Conclusions} 
In this paper, {\color{black}we have studied mirror energy differences for the $T_{z}=\pm 2$ for A=20-36 mirror nuclei using INC interactions: USDC and USDCm.} The MEDs have been extensively investigated and discussed in the mirror nuclei $^{20}$Mg/$^{20}$O, $^{22}$Al/$^{22}$F, 
$^{24}$Si/$^{24}$Ne, $^{26}$P/$^{26}$Na, $^{28}$S/$^{28}$Mg, $^{30}$Cl/$^{30}$Al, $^{32}$Ar/$^{32}$Si, $^{34}$K/$^{34}$P, and $^{36}$Ca/$^{36}$S by examining the occupation of single-particle states that are unbound or weakly bound, with an emphasis on the $1s_{1/2}$ orbital. Overall, there is good agreement between the shell model results and the experimental data. We have successfully reproduced the experimental ground state for all the mirror pairs discussed here from both the interactions, except for the $^{26}$Na ($T_Z=+2$) which is reproduced only by the USDCm interaction. The comparison between experimental and theoretical MED is shown in Fig. \ref{sd}, here considering USDC interaction $^{20}$Mg($2^+_1$), and $^{28}$S($2^+_1$) and considering USDCm interaction $^{20}$Mg($2^+_1$), $^{28}$S($2^+_1$), and $^{32}$Ar($2^+_1$) lies very close to the diagonal line. Therefore, for these mirror pairs, we have nicely reproduced experimental MED values. The calculated isoscalar ($M_0$) matrix element shows good agreement with experimentally available data for $A=20$, $28$ and $32$ mirror pairs. The calculated isovector ($M_1$) matrix element shows good agreement with experimentally available data for $A=20$, $24$, and $32$ mirror pairs. Considering the magnetic moment, there are experimental data available only for $T_z=+2$ nuclei, and results for $^{20}$O, $^{24}$Ne, $^{30}$Al, and $^{36}$S ($T_z=+2$) nuclei matches nicely or lies on the experimental error bar. The calculated single proton ($s_{1p}$) and neutron separation ($s_{1n}$) energies using the USDC interaction show very good agreement with their experimental values. Proton drip line nuclei $^{22}$Al, $^{26}$P, $^{30}$Cl, and $^{34}$K form a weakly bound system, due to very small or negative single proton separation energy.

\section*{ACKNOWLEDGMENTS}
 We are thankful for the financial assistance provided by MHRD, the Government of India, and SERB (India), CRG/2022/005167. We acknowledge the National Supercomputing Mission (NSM) for providing computing resources of ‘PARAM Ganga’ at the IIT Roorkee, implemented by C-DAC and supported by the Ministry of Electronics and Information Technology (MeitY) and Department of Science and Technology (DST), Government of India. We would like to thanks Prof. B.A. Brown for several useful discussions. We also thank Prof. V.K.B. Kota for useful suggestions.

%%%%%%%%%%%%%%%%%%%%%%%%%%%%%%%%%%%%%%%
%\section*{{References}}

\end{document}